\documentclass[floatfix,prd,lengthcheck,showpacs,amssymb,amsmath,amsfonts,aps,altaffilletter,nofootinbib,nopreprintnumbers,showpacsm]{revtex4-1}

\usepackage[utf8]{inputenc}
\usepackage[dvipsnames]{xcolor}
\usepackage{graphicx}
\usepackage{subfig}
\usepackage{amsmath}
\usepackage{acronym}
\usepackage{multirow}
\usepackage{tabu}
\usepackage{import}
\usepackage{hyperref}
\usepackage{lineno}
\usepackage{animate}
\usepackage{algorithm}
\usepackage[noend]{algpseudocode}
\hypersetup{
    colorlinks=true,
    linkcolor=blue,
    filecolor=magenta,      
    urlcolor=cyan,
}

\begin{document}
\def\thefootnote{*}\footnotetext{These authors contributed equally to this work.}\def\thefootnote{\arabic{footnote}}
%\title[]{Anomaly detection of transient noise bursts in LIGO with auto-encoders}
\title[]{Detection of anomalies amongst LIGO's glitch populations with autoencoders}
%\title[]{}
\author{Paloma Laguarta$^{1*}$}
\author{Robin van der Laag$^{2*}$}
\author{Melissa Lopez$^{2,3}$}
\author{Tom Dooney$^{4}$}
\author{Andrew L. Miller$^{2,3}$}
\author{Stefano Schmidt$^{2,3}$}
\author{Marco Cavaglia$^{5}$}
\author{Sarah Caudill$^{6, 7}$}
\author{Kurt Driessens$^{1}$}
\author{Jo\"el Karel$^{1}$}
\author{Roy Lenders$^{8}$}
\author{Chris Van Den Broeck$^{2,3}$}

\address{$^1$Department of Advanced Computing Sciences, Maastricht University, Maastricht, the Netherlands.}
\address{$^2$Institute for Gravitational and Subatomic Physics (GRASP) Utrecht University, Princetonplein 1, 3584 CC, Utrecht, the Netherlands.}
\address{$^3$Nikhef Science Park 105,
1098 XG, Amsterdam, the Netherlands.}
\affiliation{$^4$ Centre of Actionable Research Open University, Heerlen, the Netherlands}

\affiliation{$^5$Physics Department
Missouri University of Science and Technology
1315 N. Pine St., Rolla, MO 65409, USA}
\affiliation{$^6$Department of Physics, University of Massachusetts, Dartmouth, MA 02747, USA}

\affiliation{$^7$Center for Scientific Computing and Data Science Research, University of Massachusetts, Dartmouth, MA 02747, USA}
\affiliation{$^8$ Boosting Alpha B.V., Villafloraweg 1, 5928 SZ Venlo, the Netherlands}

% Please add ORCIDs here in case we submit to a journal which accepts them
% Gareth: 0000-0002-0355-5998
% Alex: 0000-0002-1850-4587
% Marton: 0000-0002-5354-5683

\date{\today}

\begin{abstract}
Gravitational-wave (GW) interferometers are able to detect a change in distance of $\sim$ 1/10,000th the size of a proton. Such sensitivity leads to large appearance rates of non-Gaussian transient noise bursts in the main detector strain, also known as glitches. These glitches come in a wide range of frequency-amplitude-time morphologies and are caused by environmental or instrumental processes, hindering searches for all sources of gravitational waves. Current approaches for their identification use supervised models to learn their morphology in the main strain, but do not consider relevant information provided by auxiliary channels that  monitor the state of the interferometers nor provide a flexible framework for novel glitch morphologies. In this work, we present an unsupervised algorithm to find anomalous glitches. We  encode a subset of auxiliary channels from LIGO Livingston in the fractal dimension, a measure for the complexity of the data, and learn the underlying distribution of the data using an auto-encoder with periodic convolutions. In this way, we uncover unknown glitch morphologies, and overlaps in time between different glitches and misclassifications. This led to the discovery of anomalies in $6.6 \%$ of the input data. The results of this investigation stress the learnable structure of auxiliary channels encoded in fractal dimension and provide a flexible framework to improve the state-of-the-art of glitch identification algorithms.
\end{abstract}

\maketitle

\section{Introduction}

{ {The first detection of a gravitational wave (GW) signal from a binary black hole (BBH) event \cite{PhysRevLett.116.061102}} by the Laser Interferometer Gravitational-Wave Observatory (LIGO) and Virgo  {collaborations} established the field of GW astronomy \cite{LIGOScientific:2014pky, VIRGO:2014yos}. {Since then, {over 90 confident astronomical events have been detected} in the past three observation runs by LIGO-Virgo collaboration \cite{LIGOScientific:2018mvr, LIGOScientific:2020ibl, LIGOScientific:2021djp} and other research groups \cite{Nitz:2019, Nitz:2020oeq, Nitz:2021uxj, Nitz:2021zwj, Zackay:2019btq, Olsen:2022pin}}. In  2017, after an improvement of the detector configuration, {the joint observation of Advanced LIGO and Advanced Virgo led to the first detection of} a binary neutron star (BNS) inspiral, labelled as GW170817 \cite{LIGOScientific:2017vwq}. The initial announcement of the detection by the Fermi Gamma-ray Burst (GRB) Monitor of GRB170817A \cite{Meegan_2009, Goldstein_2017}, and the precise sky location of GW170817 by GW detectors, enabled a rapid electromagnetic follow-up which led to the detection of the associated kilonova, later called AT2017gfo \cite{Perego:2017wtu}.} 

{The detection of GW170817 posed the added challenge of { mitigating the effect of a transient non-astrophysical burst of non-Gaussian noise from the data}, also known as a glitch, for its subsequent analysis \cite{blackburn2008lsc, abbott2016characterization}. Glitches may be caused by the environment (e.g., earthquakes, wind, anthropogenic noise) or instruments (e.g., {control systems, electronic components }\cite{LIGO:2020zwl}), though in many cases their causes remain unknown \cite{Cabero:2019orq}.  They come in a large variety of time-frequency morphologies, have a typical duration of {between sub-seconds and seconds}, {and have a high rate of occurrence }($\sim 1 $ per minute {during the first half of the third observing run, O3a} \cite{LIGOScientific:2020ibl}). They can reduce the amount of analyzable data increasing the noise floor, produce false positives in GW data, affect the estimation of the detector power spectral density and reduce candidate significance in searches for {short- and long-lived GW signals} \cite{abbott2018effects, KAGRA:2022dwb,Steltner:2023cfk, KAGRA:2021kbb, Steltner:2021qjy}.

Glitches can also bias astrophysical parameter estimation, making it difficult to determine which part of the signal corresponds to a glitch and which part to the actual GW event \cite{pankow2018mitigation,davis2019improving,driggers2019improving}. Additionally, glitches can impact line-cleaning procedures in GW searches, which rely on replacing disturbed frequency bins with artificially generated data, consistent with their neighbours \cite{Powell:2018csz, LSC:2018vzm, Steltner:2021qjy}. If the surrounding data contains elevated noise floors, the efficacy of mitigation methods will be reduced. 

{{Glitch identification and characterization is a crucial first step towards their mitigation} 
\cite{davis2020utilizing, LIGO:2021ppb}. 
{ {Most of the current approaches to glitch characterization with ML utilize supervised classification algorithms}, where models {learn to identify glitches through labelled time-frequency representations} of GW strain data $h(t)$ \cite{zevin2017gravity, George:2017fbn, bahaadini2018machine, Glanzer:2022avx, Ferreira:2022vrp, RAZZANO2023167959, alvarez2023gspynettree}. However, this procedure presents several limitations. 
{Supervised learning needs fixed class definitions that are not exhaustive nor representative of all glitch morphologies, as there could be many possible sub-classes to discover \cite{bahaadini2018machine}. Furthermore, as GW detectors are improved, novel glitch morphologies could arise \cite{Soni:2021cjy}.}
Moreover, generating these labels is an expensive task, since ML methods need a lot of examples for training{, and {experts must vet the labelling procedure}.

{In this context, unsupervised methods to identify glitches based on ML algorithms could help overcome such limitations. In this paper, we propose a novel ML algorithm that combines auxiliary channel information with an unsupervised anomaly detection algorithm. We encode the information from auxiliary channels from LIGO Livingston in the fractal dimension, a measure of the complexity of the time series. This representation of the data is input to a data-driven algorithm, which consists of a convolutional autoencoder with periodic convolutions that learns the underlying representation of the data, clustering glitches according to their similarity in a compressed representation. By exploiting this compressed representation for anomaly detection, we can identify glitches that strongly deviate from the general distribution of the input data, improving the understanding of glitch populations. We test the method's performance by identifying  {anomalies} on three classes of known glitches in LIGO data.

This paper is structured as follows. In section \ref{sec:SOTA} {we introduce the current state-of-the-art glitch characterization and explain the fractal dimension encoding. In section \ref{sec:dataset} we provide details about data acquisition and its pre-processing. In section \ref{sec:methodology} we describe the ML method employed in this investigation. In section \ref{sec:results} we present the main results of this research, showing different anomalies found with our methodology, and in section \ref{sec:conclusions} we conclude {, proposing avenues for future research}.}

\section{{Identification of detector transient noise}}\label{sec:SOTA}

\subsection{Characterization via auxiliary channels}

{The status of GW detectors is continuously monitored through a large set of data streams at various sampling rates, { outputting  $\sim 10^6$ time-series from} instrumental and environmental sensors. These auxiliary channels can be divided into safe (insensitive to GW) and unsafe (sensitive to GW). Depending on their origins, glitches present varied morphologies in different sets of auxiliary channels.  {Some subset of these channels may serve as} ``witnesses" of glitches and are used to create data quality flags before performing GW searches \cite{abbott2020guide, harry2010advanced, smith2011hierarchical}.}

{Despite {the huge amount of} auxiliary channels in a single detector, {many of them do not provide useful information for noise transient investigations} as they remain constant or vary with a consistent pattern {, constituting a data set containing redundant and/or non-informative characteristics} \cite{Colgan:2019lyo, Essick:2020qpo, abbott2016characterization}. {Therefore, LVK researchers have compiled a ``reduced" standard list of $\sim 10^3$ auxiliary channels that are used in data quality investigations. In this work, we limit our investigation to safe auxiliary channels with sampling rates $>512 \, $Hz, yielding a set of 347 channels.}}

\subsection{Fractal dimension}

{ {The first step towards  {characterizing} glitches through safe auxiliary channels requires identifying anomalous data stretches within them} %Several methods have been proposed and standard methods in LVK collaboration 
\cite{Essick:2020qpo, robinet2016omicron, smith2011hierarchical}.} {In \cite{cavaglia2022characterization}, the author proposes the measurement of fractal dimension (FD) as an additional effective tool for characterizing the instrument output in low latency. FD is an index that characterizes the self-similarity of a set and provides a measure of the complexity of the signal in the context of signal processing \cite{theiler1990estimating}. {There are several definitions of this magnitude \cite{gneiting2012estimators, tricot1982two, fernandez2014fractal}, implying that the FD measure for a physical process can differ depending on the chosen definition.} 
{Nonetheless, in this work, we focus on the \textit{FD variation over time} as an indicator of the evolution of the signal's complexity. As the presence of a glitch in the data affects the noise power spectrum, which in turn varies the value of FD, we are only interested in the relative change which is definition independent.}

%This is sufficient for our work because the chosen definition does not have a significant effect on it, since for any definition of the FD, a change in the value of the FD is associated with a change in the noise power spectrum of the data.

%Nonetheless, in this work, {we focus on the \textit{FD variation over time}, as an indicator of the evolution of the signal's complexity, which is definition-independent}, since changes in the value of the FD are associated with changes in} {the noise power spectrum of the data.}

 {To illustrate this, Fig. \ref{fig:fdpaper} presents the variation of FD for two minutes of data from the \texttt{L1:LSC-PRCL$\_$OUT$\_$DQ} auxiliary channel, which measures the Power Recycling Cavity Length (PRCL) from the Length Sensing Contol (LSC) of the LIGO Livingston (L1) interferometer. The computation was performed with a time window $\mathcal{W}(t) = 1 \,$s, i.e. every FD value is the result of encoding $1 \,$s of the input data.} { {Points greater than one standard deviation  $\sigma$ from the mean FD correlate to the presence  {of \texttt{Whistle} glitches} in the detector.}}  {As we can observe from Fig. \ref{fig:fdpaper}, FD can be an effective tool to further understand the coupling between glitches and auxiliary channels. To extend this analysis to a larger set of safe auxiliary channels and glitch classes, we first need to speed up the FD calculation to near-real time.}

\begin{figure}
    \includegraphics[width=1.\columnwidth]{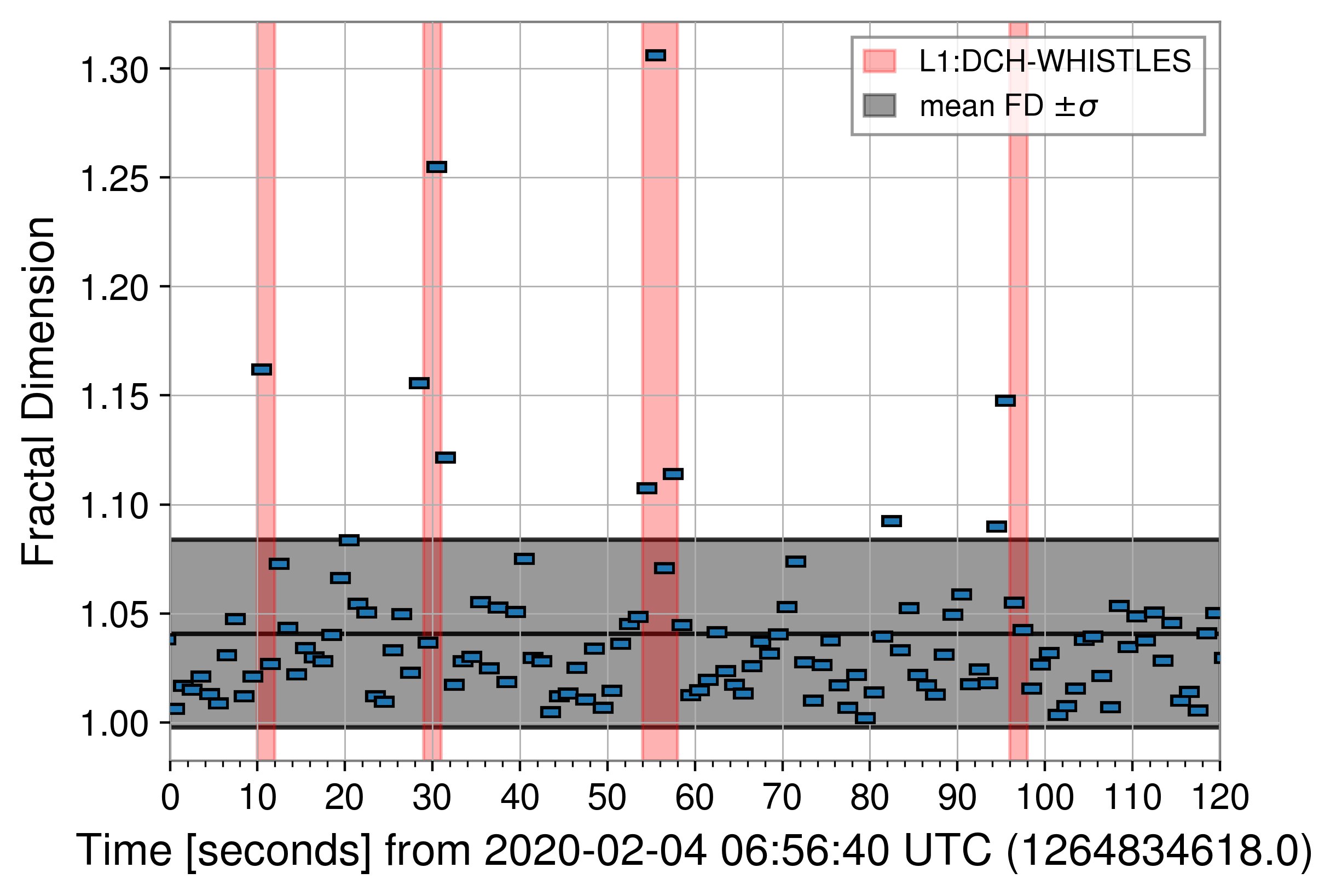}
    \caption{{Fractal dimension over a two minute period of L1 data for the \texttt{L1:LSC-PRCL$\_$OUT$\_$DQ} auxiliary channel. Each point represents the fractal dimension for one second of data and the red regions indicate the time period containing \texttt{Whistle} glitches.}}
    \label{fig:fdpaper}
\end{figure}

{Following \cite{cavaglia2022characterization} we numerically estimate the measured FD with  {the variation (VAR)} method (see \cite{cavaglia2022characterization} for details).}
 {For a discretely-sampled set of data with $N$ measurements $\mathcal{C}\in\mathbb{R}^N$}, we can define a sliding window to compute the variation of the data with centre $l$ and scale $k$,

\begin{equation}
    \mathcal{F}_{k,l}=\left| \max\left[\mathcal{C}_{l-k},\dots,\mathcal{C}_{l+k}\right] - 		\min\left[\mathcal{C}_{l-k},\dots,\mathcal{C}_{l+k}\right] \right|.
\end{equation}}

Thus, the VAR estimator for a given scale $k$ is, 
\begin{equation}
\mathcal{A}^{\textbf{VAR}}(k) = \frac{1}{N-2k}\sum_{l=k}^{N-1-k} \mathcal{F}_{k,l},
\end{equation}

{As we can see in Algorithm \ref{alg:algo_cavaglia}, the implementation in \cite{cavaglia2022characterization} computes the maximum and minimum over a range of values at each iteration $k, l$ (line 5 and 6). The runtime of this implementation is $\mathcal{O}(N^3)$.}

A significant speed-up can be achieved using Algorithm \ref{alg:algo_new} based on \cite{dubuc1996error}. It uses the fact that at iteration $k+1$ we can compute the maximum as 

\begin{equation}
\begin{split}
\max [ \mathcal{C}_{l-(k+1)}& , \dots,  \mathcal{C}_{l+(k+1)} ] = \\
 &  \max\{\max\left[\mathcal{C}_{(l-1)-k},\dots,\mathcal{C}_{(l-1)+k}\right], \\
 &  \ \qquad \max\left[\mathcal{C}_{(l+1)-k},\dots,\mathcal{C}_{(l+1)+k}\right]\},
\end{split}
\end{equation}
where the components of the right-hand side have already been computed at iteration $k$. 
This step is done on line 10 in Algorithm \ref{alg:algo_new}, and likewise in line 11 for the minimum.  {Now, the computational complexity of the FD calculation is} $\mathcal{O}(N^2\log(N))$ and the practical speed-up can be seen in Fig. \ref{fig:alg_benchmark}, where we compute FD with both methods over data increasing in length. While this speed-up is not apparent for short stretches of data at low sampling rates, it becomes significant at sampling rates $\geq 4096\,$Hz.

\begin{algorithm}[H]
    \caption{Implementation of the VAR method from \cite{cavaglia2022characterization}.}\label{alg:algo_cavaglia}
    \textbf{Input:}\quad\ $f$ vector of size $N$.\\
    \textbf{Output:}\quad $A$ vector of size $N/2$.
    \begin{algorithmic}[1]
        \State $A[1\dots N/2]=0$
        \For {$k=1$ \textbf{to} $N/2$}
            \State $F = \emptyset$
            \For {$l=k$ \textbf{to} $N-k$}
                \State $F = F \cup \{\max\{f[l-k],\dots,f[l+k]\} $
			\State \phantom{$F = F \cup \{$}$- \min\{f[l-k],\dots,f[l+k]\}\}$
            \EndFor
            \State $A[k] = \text{mean}(F)$
        \EndFor
        \State \textbf{return} $A$
    \end{algorithmic}
\end{algorithm}

\begin{algorithm}[H]
	\caption{Improved algorithm for the VAR method.}\label{alg:algo_new}
	\textbf{Input:}\quad\ $f$ vector of size $N$.\\
	\textbf{Output:}\quad $A$ vector of size $N/2$.
	\begin{algorithmic}[1]
		\State $A[1\dots N/2]=0$
		\State $u[1\dots N-2]=0$
		\State $b[1\dots N-2]=0$
		\For {$i=1$ \textbf{to} $N-2$}
		\State $u[i] = \max\{f[i],\dots,f[i+2]\}$
		\State $b[i] = \min\{f[i],\dots,f[i+2]\}$
		\EndFor
		\State $A[1] = \text{mean}(u[1\dots N-2]-b[1\dots N-2])$
		\For {$i=2$ \textbf{to} $N/2$}
		\For {$j=1$ \textbf{to} $N-2i$}
		\State $u[j] = \max\{u[j], u[j+2]\}$
		\State $b[j] = \min\{b[j], b[j+2]\}$
		\EndFor
		\State $A[i] = \text{mean}(u[1\dots N-2i]-b[1\dots N-2i])$
		\EndFor
		\State \textbf{return} $A$
	\end{algorithmic}
\end{algorithm}

\begin{figure}
    \includegraphics[width=1.\columnwidth]{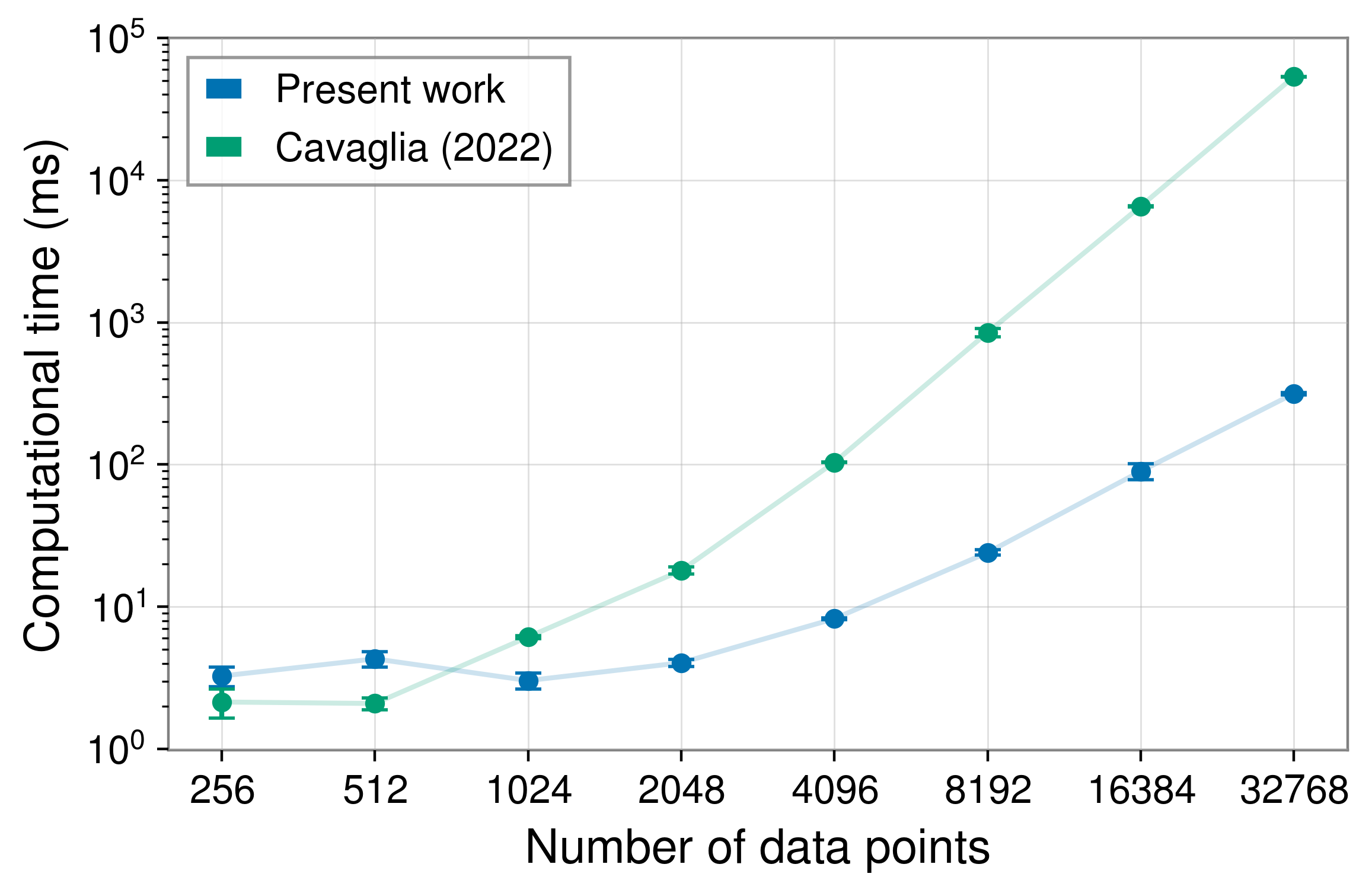}
    \caption{{Comparison of the fractal dimension  computing algorithms for varying number of data points. Benchmarks done on a Intel\textsuperscript{\tiny\textregistered} Xeon\textsuperscript{\tiny\textregistered} Processor E5-2630 v4 CPU @ 2.20GHz.}}
    \label{fig:alg_benchmark}
\end{figure}

 {In practice, with Algorithm \ref{alg:algo_cavaglia} with computational complextity  $\mathcal{O}(N^3)$, we were able to FD-encode $1 \,$h of data in $1 \,$h, but now with an efficient implementation with \texttt{numba} \cite{lam2015numba} of Algorithm \ref{alg:algo_new} based on \cite{dubuc1996error}, with computational complextiy $\mathcal{O}(N^2\log(N))$, we can now process $1 \,$h in $11 \,$s. With further parallelization in a cluster, the FD computation could characterize glitches in low latency. Now that we have a fast computation of FD-value, we can construct a data set for our application.}

\section{Data set and pre-processing}
\label{sec:dataset}

 {The aim of this work is to understand the underlying glitch population using solely information from safe auxiliary channels. Due to the overwhelming amount of information, we encode the data of the safe auxiliary channels in FD. Afterwards, we use an unsupervised ML method to learn the underlying distribution of the data, finding anomalies that strongly deviate from the general trend of the FD-encoded data. While unsupervised ML algorithms in the context of anomaly detection are agnostic, as they do not make prior assumptions regarding the data distribution, it is challenging to interpret their results. To understand the results of our algorithm and assess its performance we can compare the output of our algorithm with the findings of supervised glitch classifiers, employing them as a benchmark. In the following subsections, we describe the benchmark used in this work, the selection of glitch populations and the FD-encoding of the data.}

\subsection{Glitch classification}

In the present work, we employ \textit{Gravity Spy} as a benchmark, finding anomalies from its high-confidence classifications. \textit{Gravity Spy} is an algorithm that combines supervised ML  and citizen science to characterize glitches present in LIGO data according to their morphologies in GW strain data $h(t)$, {represented in time-frequency}  \cite{zevin2017gravity} . The trained algorithm assigns glitches a pre-defined class and gives a confidence score that it belongs to this class.} 

{In practice, alerts are generated by Omicron, which is an algorithm designed to search for power excess in time series data using the Q-transform, a modification of the standard short-time Fourier transform parameterized by a quality factor Q \cite{brown1991calculation, robinet2016omicron}.  {Gravity Spy assigns a class and a confidence value to Omicron's alert if it exceeds $7.5\,$signal-to-noise ratio (SNR), a magnitud related to the tranform coefficient of the Q-transform.} Currently, Gravity Spy considers 22 glitch classes, which have been previously identified \cite{aasi2015characterization, abbott2016characterization, nuttall2015improving}.}

\subsection{Glitches}
\label{sec:morphologies}

 {In this proof-of-concept work, we select GPS times $t$ that contain in $h(t)$ no apparent excess of power, and $t$ of three distinct glitch morphologies in LIGO Livingston with Gravity Spy confidence $> 90 \%$ \cite{Glanzer:2022avx}. One must note that for the glitches $t$ represents the peak time of the Omicron alert. The three morphologies are chosen to have short and long-duration glitches that are abundant in LIGO Livingston data ($> 800$ samples per class), and that impact GW searches due to their wide frequency contribution. We detail each class below:}

\begin{itemize}
\item \texttt{No$\_$Glitch}: in this class, no significant excess power is visible in the Gravity Spy spectrograms (see Fig. \ref{fig:mclean}). In the context of this work, this class represents a stable behaviour of the GW detector, which is reflected by non-deviant FD values. 

\item \texttt{Whistle}: these glitches have a characteristic V, U or W shape at higher frequencies ($\gtrsim$ 128 Hz) with typical durations $\sim 0.25\,$s.  {They are caused when radio-frequency signals beat with the voltage controlled oscillators \cite{Nuttall:2015dqa}.} In Fig. \ref{fig:mwhistle} we present a \texttt{Whistle} glitch with a frequency content $> 512\,$Hz.

\item \texttt{Tomte}: these glitches are also short-duration ($\sim 0.25\,$s) with a characteristic triangular morphology. In  Fig. \ref{fig:mtomte} we show a \texttt{Tomte} glitch from LIGO Livingston, where these morphologies are quite abundant. Since there is no clear correlation to the auxiliary channels, they cannot be removed from astrophysical searches.

\item \texttt{Scattered$\_$Light}: also known as Slow Scattering, these glitches have longer duration harmonics ($\sim2.0 – 4.0\,$s) that in time-frequency domain they appear as arches being often stacked on top of each other (see Fig. \ref{fig:mscatlig}). These glitches are quite problematic since their frequency content lies in the band of interest of GW astrophysical events.  {In O3, they were found to be coupled with the relative motion between the optical suspension system’s end test-mass chain and the reaction-mass chain \cite{Soni:2021}}.
\end{itemize}

In this work we focus on LIGO Livingston, as the author in \cite{cavaglia2022characterization}, but this investigation could be extended to LIGO Hanford and Virgo. {The details on how this data was pre-processed for its posterior usage in our model, can be found in the next section.}

\begin{figure*}
\subfloat[\label{fig:mclean}{\texttt{No$\_$Glitch} class}]{%
\includegraphics[width=0.572\columnwidth]{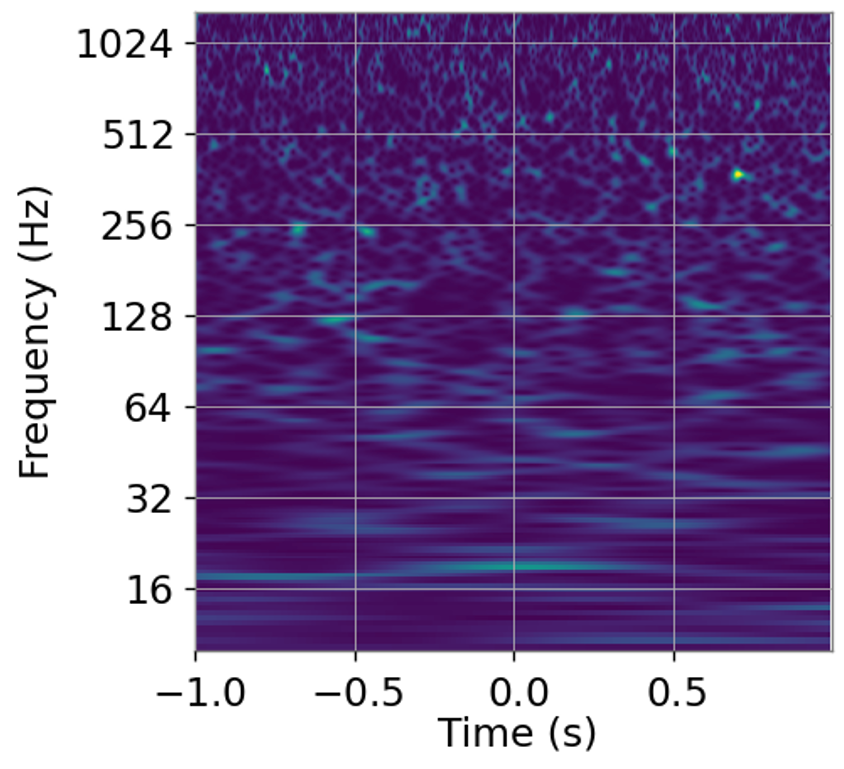}%
}\hfill
\subfloat[\label{fig:mwhistle}{\texttt{Whistle} class}]{%
\includegraphics[width=0.48\columnwidth]{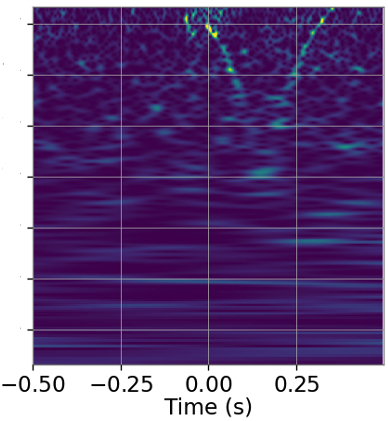}%
}\hfill
\subfloat[\label{fig:mtomte}{\texttt{Tomte} class}]{%
\includegraphics[width=0.46\columnwidth]{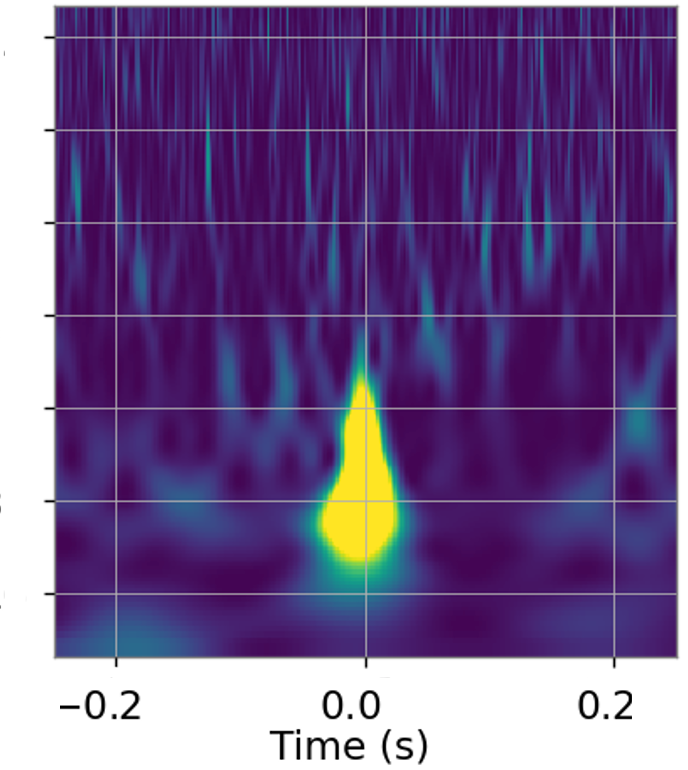}%
}\hfill
\subfloat[\label{fig:mscatlig}{\texttt{Scattered$\_$Light} class}]{%
\includegraphics[width=0.56\columnwidth]{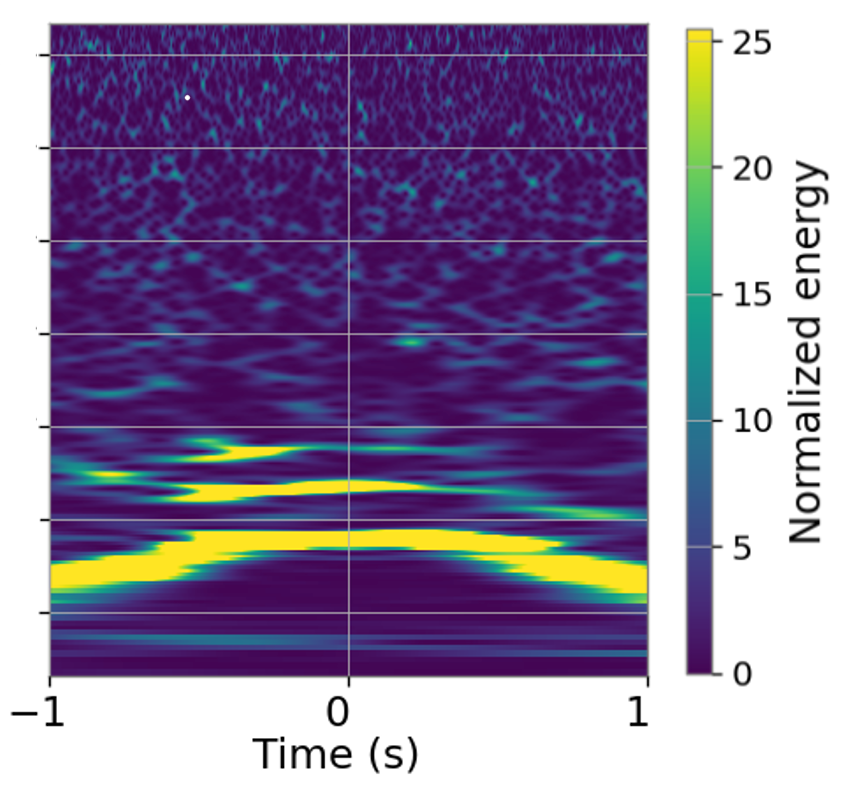}%
}
\caption{From left to right: Q-transform of \texttt{No$\_$Glitch}, \texttt{Whistle}, \texttt{Tomte} and \texttt{Scattered$\_$Light} retrieved from Gravity Spy \cite{zevin2017gravity}.  }
\label{fig:morphologies}
\end{figure*}

\subsection{Auxiliary channels encoded in fractal dimension}\label{sec:reduced}

 { {Given a time $t$ of interest}, we select an array of GPS time with duration {$\Delta t = 8$}s, where $t$ is in the center. For each array of time, we retrieve $347$ safe auxiliary channels with sampling rates $> 512\,$Hz, excluding the GW strain $h(t)$, that is then whitened and encoded in FD with time windows $\mathcal{W}(t) \in \{0.25, 0.5, 1, 2\}\,\mathrm{s}$.} {For each $\mathcal{W}(t)$ we have $\Delta t/\mathcal{W}(t) -1$ time bins to ensure that $t$ is in the center of the FD-encoded data,   {yielding} a total of 56 time-bins for sample.} Since  {the duration of \texttt{Scattered$\_$Light} is $\sim 2 \-- 4\,$s and the duration of \texttt{Tomte} is $\sim 0.25\,$s},  {the length of these} varying time windows ensure that any glitch morphology will be contained {at least within $\mathcal{W}(t) =2$s} \cite{zevin2017gravity}. Note that the sampling rate of each independent auxiliary channel varies, but we only encode safe channels with a sampling rate $> 512\,$Hz, to have enough data points to perform a {calculation of the FD}, { as demonstrated by the experiments performed by \cite{cavaglia2022characterization}}.}

{{Limited by the number of \texttt{Whistle} present in LIGO Livingston, for the initial data set we select 896 GPS times for each class defined} in Section \ref{sec:morphologies} and presented in Fig. \ref{fig:morphologies}, yielding a balanced data set. Since each auxiliary channel monitors distinct physical processes{, their average FD measurements can differ, giving priority to certain channels over others. To improve the stability of our model, we normalize in the range} $[0, 1]$ the data  {of } each individual auxiliary channel, as we are only interested in their relative variation. Normalizing collectively would give more importance to the channels with a higher FD and dismiss the channels with a lower FD.}

{ For this work we reduce the dimensionality of the normalized data set with dimensions $347 \text{ channels }\times 56 \text{ time bins }$ using a data-driven approach. Our aim is to maintain the channels that capture the most relevant features of the glitches with respect to the  \texttt{No$\_$Glitch} class, so we follow the procedure below:}

\begin{enumerate}
\item {Defining $D_{\mathrm{NG}}$  {as the set of} \texttt{No$\_$Glitch} FD-encoded, we compute the average of all elements in $D_{\mathrm{NG}}$, $\mu{(D_{\mathrm{NG}})}$,  to minimize extreme deviations of FD. This will be the common background when \texttt{No$\_$Glitch} is present.} 
\item {For a single glitch $d_{\mathrm{C}}$ encoded in from a certain class C, we subtract the background as $d_{\mathrm{C}} - \mu{(D_{\mathrm{NG}})}$. This subtraction highlights the deviations produced by the presence of a glitch in the data.}
\item  {We identify auxiliary channels $A$ that present a low FD deviation with respect to the background, since their contribution is similar to the absence of glitch.} Thus, given a glitch $d_{\mathrm{C}}$ and a channel $A$, if ${d_{\mathrm{C, A}} -\mu{(D_{\mathrm{NG}})}_{{\mathrm{A}}} \lesssim 10^{-2} \quad \forall \text{ glitches }d_{{\mathrm{C}}}}$, the auxiliary channel $A$ is removed.  {This threshold represents a balance between data compactness and expressiveness. Too many channels can introduce irrelevant information, while too few may overlook the overall data trends.}
\end{enumerate}

%{\plg{The threshold $10^{-2}$ presents a fair trade-off between discarding too many channels, which results in a representation that is too compact and cannot capture all the relevant details in the data, and keeping too many channels, which keeps irrelevant information that is similar to the FD behaviours found in the\texttt{No$\_$Glitch} class. Taking into account the interval of the FD values, the threshold $10^{-2}$ has a low enough value to be considered as 0}. 

This pre-processing reduced the dimensionality to a shape of $50 \text{ channels}\times 56 \text{ time bins }$. In Fig. \ref{fig:fdexample} we show an example of 8s of FD-encoded data. To train the ML model presented in the next section we will use the three glitch morphologies, with a total of 2688 samples, which contain the structure that we wish to unravel. {One must note that while supervised approaches must use a subset of the data to assess the generalization ability of the model, in the present unsupervised approach we are interested in learning the details of the data at hand, such that all data instances are employed.}

%it is typical to use a subset of data samples for preprocessing and training in order to not overfit the data and allow the model to gain generalisation power. In our case, as we are studying glitches through an unsupervised pipeline, there is no need for generalisation, as we want the model to get to know the data as well as possible. For this reason, all of the data instances are used for both the preprocessing and the model's training.

\begin{figure*}
    \centering
    \includegraphics[width=\textwidth]{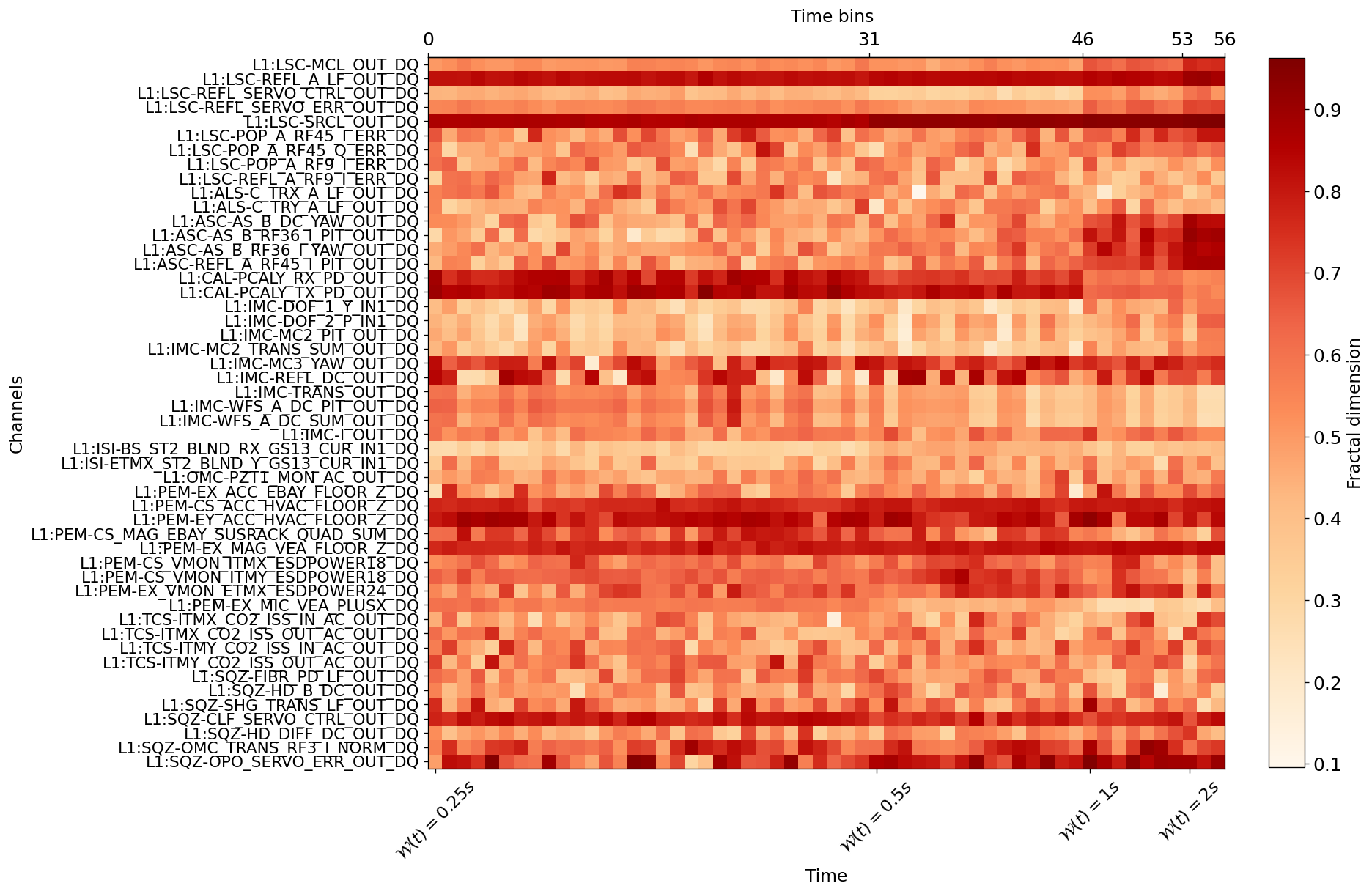}
    \caption{A sample encoded in fractal dimension with 347 safe auxiliary channels and 56 time bins.  Notably, the same event is replicated four times with a different time window: time bins $\in [0, 31)$ have a time window $\mathcal{W}(t) = 0.25\,$s, time bins $\in [31, 46)$ have $\mathcal{W}(t) = 0.5\,$s, time bins $\in [46, 53)$ have $\mathcal{W}(t) = 1\,$s, time bins $\in [53, 56]$ have $\mathcal{W}(t) = 2\,$s.}
    \label{fig:fdexample}
\end{figure*}

\section{Methodology}
\label{sec:methodology}

{In recent times, ML algorithms have sparked the interest of scientists due to their success in solving various tasks in different domains and their transversal applications. In particular, they have emerged as a novel tool in the field of GW for different tasks: pattern recognition, such as identification of BBH \cite{George:2016hay,Gabbard:2017lja}, BNS \cite{Menendez-Vazquez:2020khz, Baltus:2021nme, Baltus:2022pep}, transient burst GW  \cite{Skliris:2020qax, Lopez:2021rci, LopezPortilla:2020odz, Boudart:2022apz, Boudart:2022xib, Bini:2023gil, Meijer:2023yhn, 2020arXiv200204591C, Antelis:2021qak}, and continuous wave signals \cite{Modafferi:2023nzt}; GW signal and glitch generation \cite{McGinn:2021jqg, Lopez:2022lkd, Dooney:2022arh, Yan:2022spw, Powell:2022pcg, Lopez:2022dho}; as well as anomaly detection \cite{Morawski:2021kxv}, among others (see \cite{Cuoco:2020ogp} for a comprehensive review).} 

{The novelty of this work lies in combining auxiliary channel information with ML in the context of anomaly detection. The complexity of the dataset presented in Section \ref{sec:SOTA} implies two main challenges:}

\begin{itemize}
\item {\textit{Lack of ground-truth:} While glitch morphologies have been widely studied, there is no guarantee that all glitches belonging to a certain class present the same standard behaviour.  In this work, the labels assigned by Gravity Spy are not considered ground truth and are used only for analysis and comparison purposes. }
\item {\textit{Lack of absolute ordering:} The ordering in the channels is arbitrary, as they measure different physical magnitudes. Consequently, it is not expected to find local patterns in the vertical axis with any particular channel ordering. Thus, our model needs to learn patterns beyond local correlations in an order-independent way.}
\end{itemize}

{In the following subsections we provide the details of our implementation and how these issues were addressed.}

\subsection{Convolutional Autoencoder}

\begin{figure*}
    \centering
    \includegraphics[width=1\textwidth]{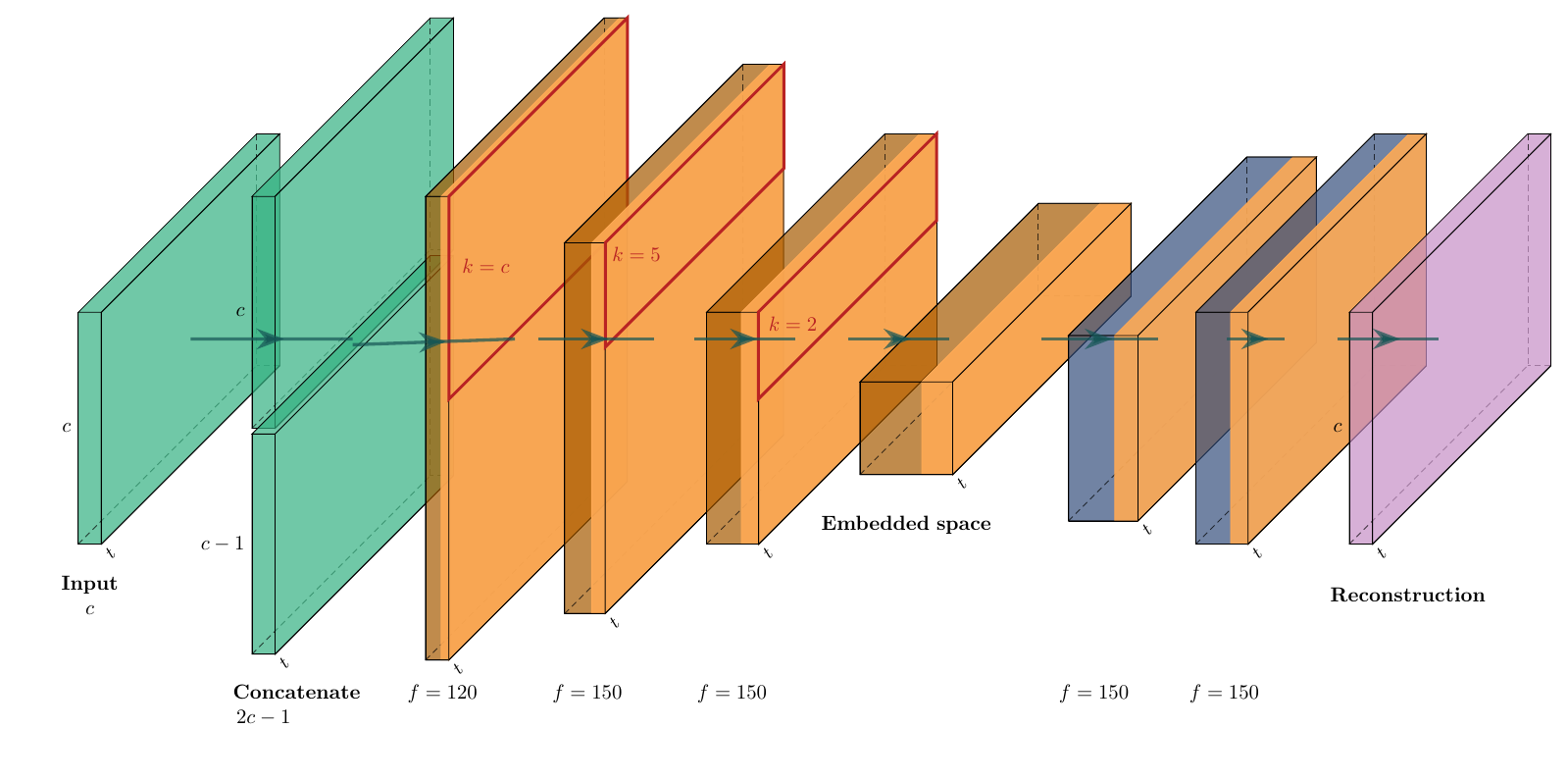}
    \caption{Outline of the autoencoder with periodic convolutions implementation. The input data which is concatenated yields a shape $(t, 2c -1)$ is presented in green. Downsampling convolutions are coloured in dark orange while upsampling convolutions are coloured in dark blue. Each convolution is followed by a ReLU activation function coloured in orange. This architecture yields a reconstruction of shape $(t, c)$.}
    \label{fig:model}
\end{figure*}

{To address the concern of lack of ground truth, we employ an autoencoder in the context of anomaly detection.}
{Autoencoders are a type of deep-learning algorithm known for their ability to uncover essential structures and patterns within unlabeled datasets, as well as their effectiveness in anomaly detection \cite{bank2020autoencoders, 2015arXiv151108458O}. They achieve this by compressing the data into a lower-dimensional or sparse format, known as an {embedded space}, maintaining the most relevant information from the dataset (encoding), and subsequently reconstructing it (decoding). The encoding is expected to employ important sub-structures that can be difficult to notice in the original representation space due to its higher dimensionality and feature redundancy \cite{2015arXiv151108458O}.}
{As a consequence, the learned embedding serves as a reference for detecting irregularities in the glitch data, since data points that deviate significantly from their embedded representations are likely anomalous. Still, the embedded space can be hard to interpret itself, since it forms a high-dimensional space in which the data is densely packed.}

{When dealing with data with a natural order between the features, convolutional filters can be used to detect (hierarchically combined) local structures and patterns that allow a complex encoding model without an exponential increase of its parameters \cite{guo2017deep, o2015introduction, zhang2018better}.  While our data is two-dimensional ({time bins} $\times$ {auxiliary channels}), given that we want to preserve the detail in the time dimension, we allow our model only to convolve along the channel dimension using 1-dimensional convolutions.}

\subsection{Periodic Convolutions}

{To address the concern of lack of absolute ordering and the possible lack of local patterns to exploit with limited range convolutional filters, we employ a periodic convolution with filters sized to cover all channels instead. We take inspiration from circular convolutions, which are used in the field of signal processing and consider the input signals as circular, or periodic, rather than finite, i.e. the end of the signal wraps around to the beginning, creating a cyclic or periodic nature \cite{priemer1991introductory}.}
{In the context of this work, we use convolutional filters with a size equal to the number of channels, so that the filter has the opportunity to ignore the data's arbitrarily chosen channel ordering.  The model applies these filters periodicly, hence the name, so that learned filters can still be used by the model to encode structures and patterns found on different channel-combinations.}

{The outline of the autoencoder's structure is presented in Fig. \ref{fig:model}. Periodic convolutions were implemented with \texttt{tensorflow} \cite{abadi2016tensorflow} and \texttt{keras} \cite{chollet2015keras} using a custom approach: given input with $c$ auxiliary channels and $t$ time bins, the input gets duplicated and concatenated along the channel axis, removing the last channel, as it is represented in Fig \ref{fig:periodic}. {In this way, each  {cyclic permutation} of channels is seen only once.}
Thus, the dimensionality of the input fed into the model is $(c, t) + (c - 1, t) = (2c -1, t)$,  {(see second block of Fig. \ref{fig:model} where the concatenation happens)}. To maintain the time dimension, we convolve along the channel dimension with a kernel size $k = c \times 1$, stride $s= 1$, filters $f = 120$ and no padding.  This custom approach ensures that there is no specific spacial ordering in the vertical (channel) dimension and that the model can capture correlations between all the channels.}

\begin{figure}
    \centering
    \includegraphics[width=1\columnwidth]{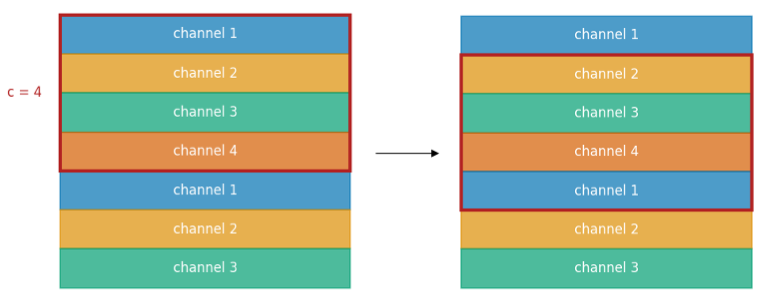}
    \caption{Example of periodic convolution where input of shape $(2c - 1, t)$ is convolved with a kernel $k=c=4$. All  {cyclic channel permutations} are convolved, so there is no absolute ordering.}
    \label{fig:periodic}
\end{figure}

{Once the first convolutional layer has captured the data patterns between arbitrary channels, the rest of the encoder architecture has the goal of constraining the data into an embedded space. Thus, it consists of two conventional downsampling convolutional layers with kernel sizes $k = 5 \times 1$ and  $k = 2 \times 1$ with $f = 150$, respectively, represented in Fig.~\ref{fig:model} in dark orange. The decoder structure is a mirror of the encoder, but with up-sampling convolutional layers instead, coloured in blue in Fig.~\ref{fig:model}. The resulting embedded space has a dimensionality of $5 \times 56$ with $f= 150$, which we will use to detect anomalies.}

{After each convolutional layer a ReLU activation function is employed to introduce non-linearity in the model (see Fig. \ref{fig:model} in orange).  The ReLU activation function avoids vanishing gradients \cite{arora2016understanding}. The model was trained for 500 epochs and a batch size of 168, using the Adam optimizer \cite{kingma2014adam}, with a learning rate $l_r = 10^{-3}$. The loss function employed is the Mean Squared Error (MSE) loss, which represents the cumulative squared error between the input and its reconstruction \cite{kim2021comparing}. To assess the performance of the model, we use the reconstruction error $\epsilon_R$ which is defined as follows:}

\begin{equation}
\epsilon_R = \sum^{c -1 }_{k=0} \sum^{t-1}_{l=0} (D_{i, kl} - D_{r, kl})^2,
\end{equation}
{where $D_{i}$ is the input to the model and  $D_{r}$ its reconstruction \cite{chai2014root}. We expect that lower reconstruction errors translate to more accurate anomaly identification.}

\subsection{t-Distributed Stochastic Neighbour Embedding}

{As discussed in the previous section, the embedded space is still high dimensional and difficult to interpret. Hence, to further lower the data dimensionality and make the data distribution in the embedded space easier to interpret, we use the t-distributed stochastic neighbour embedding (t-SNE) method\footnote{\texttt{TSNE} function from \texttt{scikit-learn} library \cite{pedregosa2011scikit} was employed.}, which projects high-dimensional data in a low-dimensional space, preserving local relationships between data points and underlying structure of the data, but releasing global relationships between data points  \cite{van2008visualizing}. While the t-SNE variables do not have an interpretable physical meaning, they are a linear correlation of physical variables and they can be traced back to assess their contribution. }

{After obtaining the 2-dimensional projection of the embedded space with t-SNE, it is straightforward to visualize the distribution of the different data points, which correspond to the glitch instances, as we show in the next section (see Fig. \ref{fig:tsnes}). The 2-dimensional plot is expected to reveal different clusters and interesting structures in the data since data points that are distant from the main clusters of their predicted class are anomalies. By labelling each glitch with its corresponding Gravity Spy label with confidence $> 90 \%$, outliers and new glitch morphologies are expected to be identified.}

\section{Results}
\label{sec:results}

{To assess the performance of the autoencoder presented in the previous section, we show  {the achieved} reconstruction errors $\epsilon_R \in [0.001, 0.014]$, as seen in Fig. \ref{fig:recerr}. The three glitch classes present similar distributions, with the \texttt{Scattered$\_$Light} class reaching the highest reconstruction error $\epsilon_R = 0.014$.  The reconstructed input differs from its original on $1.75 \%$ and $17.5 \%$ in the best and worst reconstructed pixels, respectively.  Note that the model reconstructs 98.8 $\%$ of glitches with $\epsilon_R < 0.002$, which is a low error given the range $[0.0, 1.0]$ of the input data.   In general, it appears that pixels from the same auxiliary channel have similar reconstruction errors, which translates to the preservation of  {FD-encoded data structure within the given auxiliary channel}.}

{In Fig. \ref{fig:tsnes} we present joint and marginal distributions  of the t-SNE projections with three different representations of the data: the original dataset with 347 auxiliary channels, the reduced dataset with 50 auxiliary channels (see Section \ref{sec:reduced}), and the embedded space with shape $5 \times 56$ with $f= 150$. {These t-SNE representations cluster the input data in different regions of the space, such that the samples present in the out-skirts will be considered anomalies. We use the labels of Gravity Spy to track in which regions of t-SNE space the different classes fall.} } 

\begin{figure}[h]
    \centering
    \includegraphics[width=1\columnwidth]{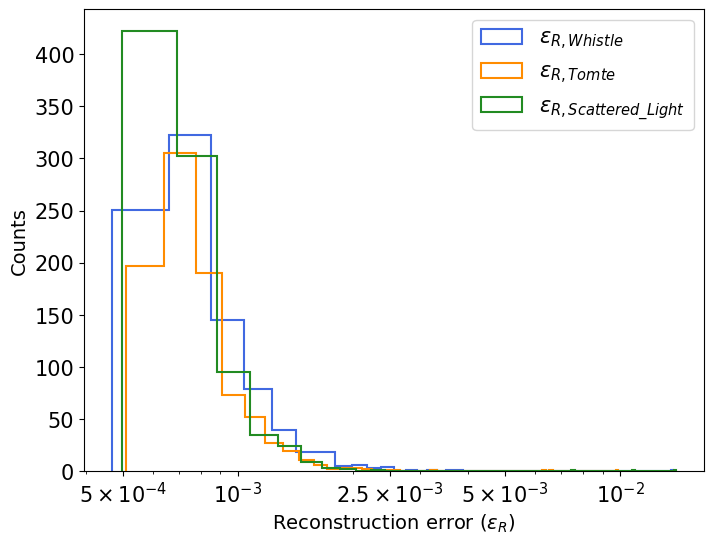}
    \caption{Histogram with the reconstruction errors $\epsilon_{R}$ for each glitch class in logarithmic scale.}
    \label{fig:recerr}
\end{figure}

\begin{figure}[!]
\subfloat[\label{fig:originaltsne}{\textit{Original t-SNE:} projection of FD-encoded data with 347 auxiliary channels.}]{%
\includegraphics[width=1\columnwidth]{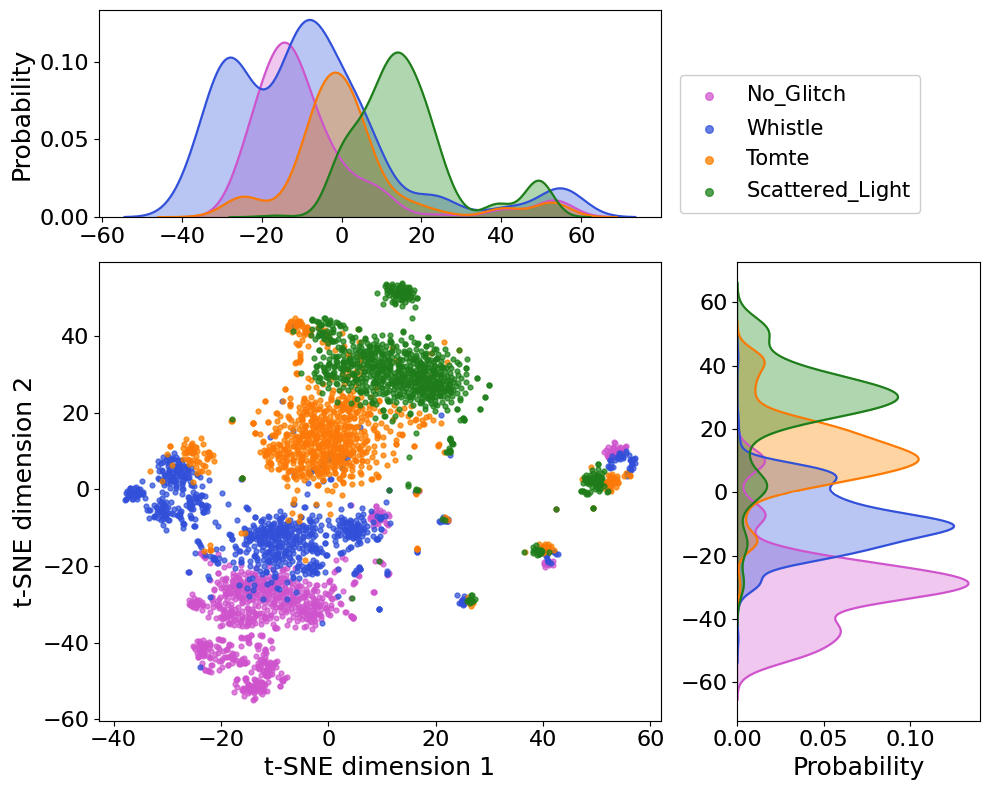}%
}\hfill
\subfloat[\label{fig:reducedtsne}{\textit{Reduced t-SNE:} projection of FD-encoded data with 50 auxiliary channels}]{%
\includegraphics[width=1\columnwidth]{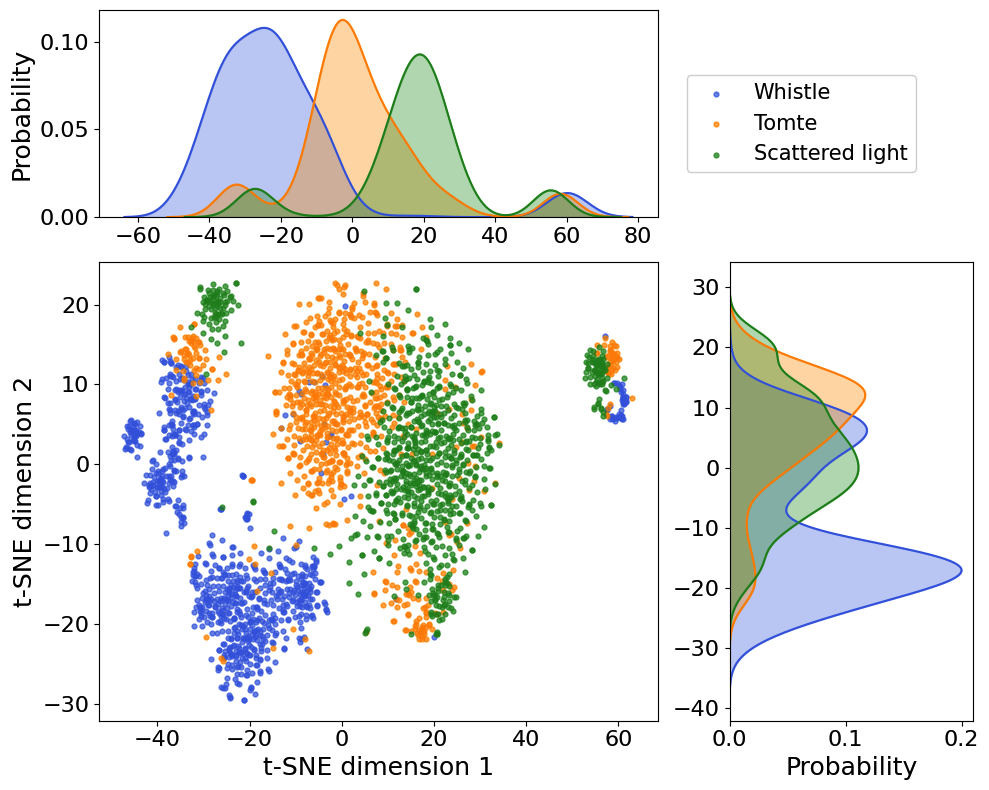}%
}\hfill
\subfloat[\label{fig:embeddedtsne}{\textit{Embedded t-SNE}: projection of the embedded space of the autoencoder showing five regions of interest.}]{%
\includegraphics[width=1\columnwidth]{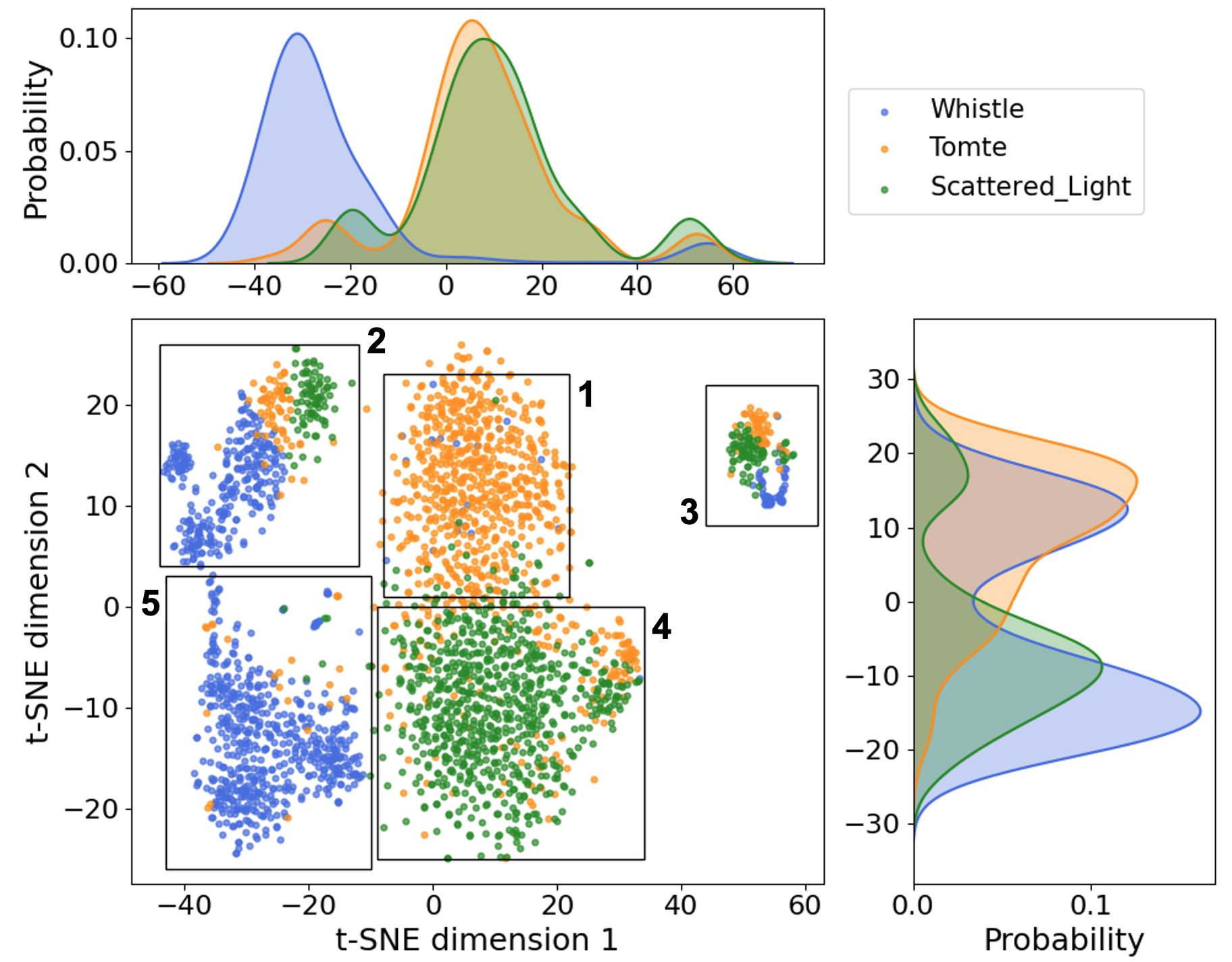}%
}
\caption{{Joint and marginal distributions of t-SNE projections of the data.}}
\label{fig:tsnes}
\end{figure}

{In Fig \ref{fig:originaltsne}, the t-SNE projection of the original FD-encoded data with 347 safe auxiliary channels  {clusters the glitch by similarity, which is consistent with  Gravity Spy's classification revealing} some overlap between classes, especially between \texttt{Tomte} and Scattered$\_$Light, as well as between \texttt{Whistle} and \texttt{No$\_$Glitch}. A dataset with instances of dimensions $347 \times 56$ introduces an enormous complexity. Therefore, we reduce the dimensionality of the data, }{ {using the safe auxiliary channels} that show the most variance in the FD-encoding, which is related to the presence of glitches (see Section \ref{sec:reduced}). Such reduction in dimensionality yields a more compressed representation of the input data  {and less overlap between the glitch classes}, as can be seen in Fig. \ref{fig:reducedtsne} since there are fewer sub-clusters of each class. After training the autoencoder, which yields small reconstruction errors $\epsilon_R$, we can project its embedded space in 2 dimensions with t-SNE, as we can see in Fig. \ref{fig:embeddedtsne}. We can observe that it looks similar to the reduced t-SNE in Fig. \ref{fig:reducedtsne}, but its marginal distributions seem similar to Gaussian distributions since the model is learning the general trend of local and global correlations of the FD-encoded data.
 {While at each compression we are discarding some characteristics of the data, we are maintaining the general trend of the data which, as stated before, is consistent with what is observed by Gravity Spy in $h(t)$, the main strain of the data.}}

{To explore the embedded space, in Fig. \ref{fig:embeddedtsne},  {we have manually outlined the distinct clusters}. In the following, we present the time-frequency of the strain $h(t)$ of some anomalous examples found in the outskirts of these clusters solely employing safe auxiliary channels.}  {The reader will encounter glitch classes that have not been mentioned before in the present work, but their description can be found in \cite{bahaadini2018machine}.}

\begin{itemize}
    \item \textbf{Region 1} corresponds to the main \texttt{Tomte} cluster. However, some \texttt{Whistle} and \texttt{Scattered$\_$Light} labels are also present. From this region, in  Fig.~\ref{fig:whistle_sec1} we show a glitch classified as \texttt{Whistle} but with an anomalous morphology, and in Fig.~\ref{fig:whistle2_sec1} we show a misclassified Wandering Line glitch classified as a Whistle. In Fig.~\ref{fig:sclight_sec1} we present a glitch classified as \texttt{Scattered$\_$Light} but its morphology is similar to Scratchy, and in Fig.~\ref{fig:sclight2_sec1} we can see a \texttt{Scattered$\_$Light} with an anomalous morphology.
\end{itemize}

\begin{itemize}
    \item \textbf{Region 2} is a sub-cluster of Whistle, with \texttt{Tomte} and \texttt{Scattered$\_$Light} overlaps.  In Fig.~\ref{fig:tomte_sec2}, we can see a \texttt{Tomte} glitch that is revealed to be overlapping with a Scratchy glitch in a longer time window. In Fig.~\ref{fig:tomte2_sec2} we see a \texttt{Tomte} glitch overlapping with a smaller \texttt{Tomte} glitch. Fig.~\ref{fig:sclight_sec2} shows a Fast Scattering glitch mislabelled as \texttt{Scattered$\_$Light} and Fig.~\ref{fig:sclight2_sec2} presents a \texttt{Scattered$\_$Light} overlapping with an unknown morphology.
\end{itemize}

\begin{figure*}[h]
\subfloat[\label{fig:whistle_sec1}{Glitch labeled as \texttt{Whistle}, but appears to be an anomalous morphology.}]{%
\includegraphics[width=1\columnwidth]{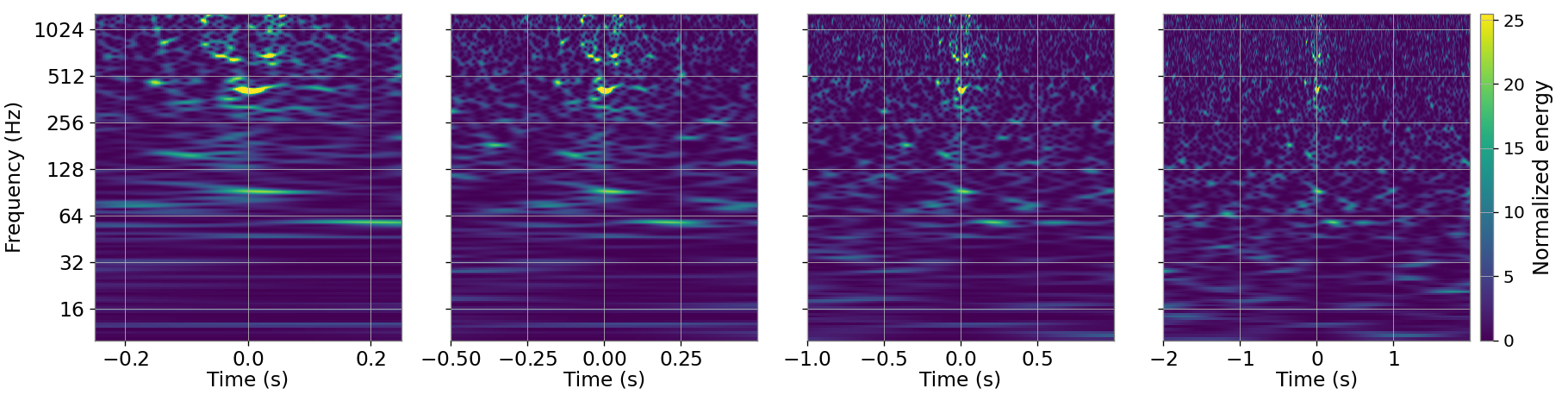}%
}\hfill
\subfloat[\label{fig:sclight_sec1}{Glitch labeled as \texttt{Scattered$\_$Light}, but its shape is consistent with the \texttt{Scratchy} class, being a misclassification. }]{%
\includegraphics[width=1\columnwidth]{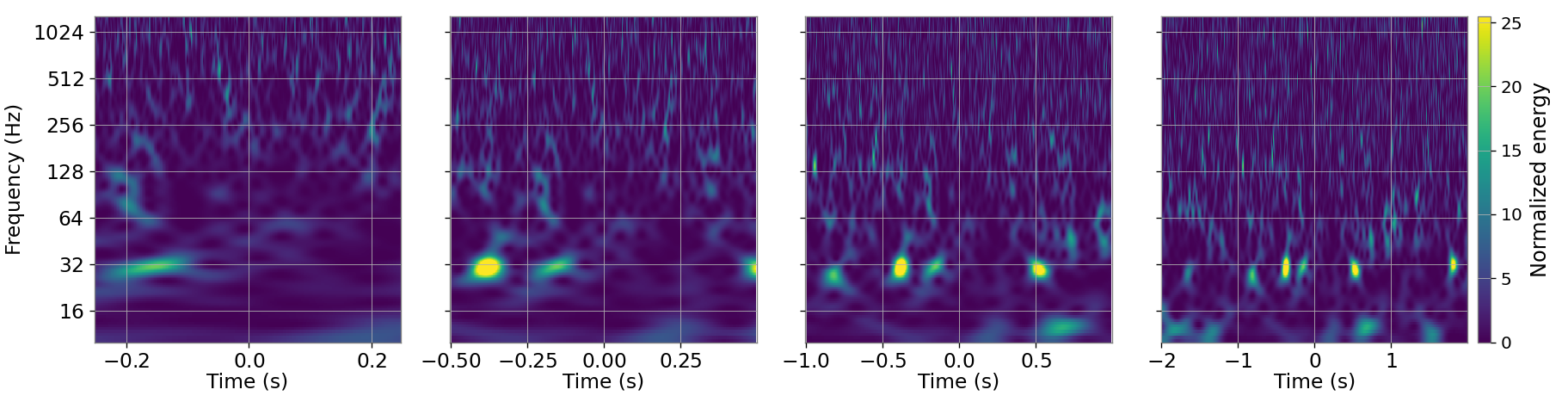}%
}\vfill
\subfloat[\label{fig:whistle2_sec1}{Glitch labeled as \texttt{Whistle}, but its shape is more consistent with \texttt{Wandering$\_$Line} class, constituting a misclassification. }]{%
\includegraphics[width=1\columnwidth]{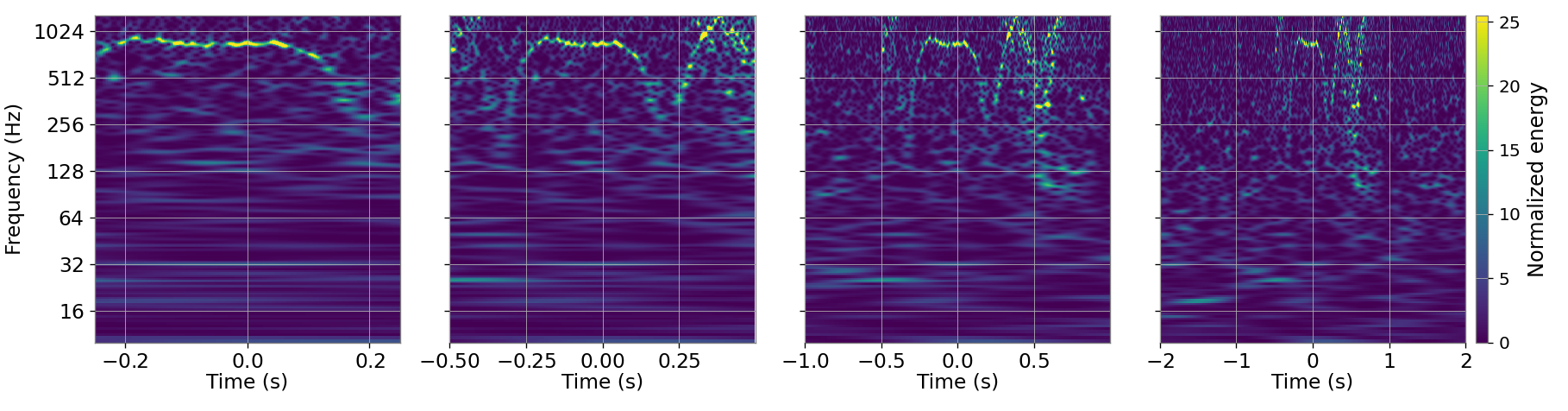}%
}\hfill
\subfloat[\label{fig:sclight2_sec1}{Glitch labeled as \texttt{Scattered$\_$Light}, but appears to be an anomalous morphology. }]{%
\includegraphics[width=1\columnwidth]{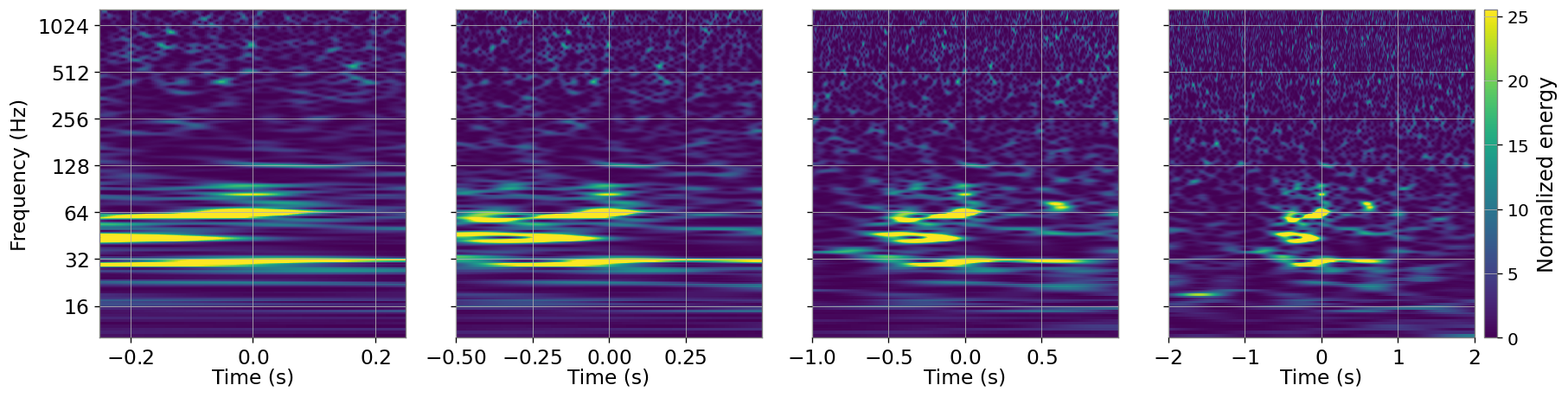}%
}
\caption{\textbf{Region 1:} Example of anomalous glitches in the embedded space.}
\label{fig:section1}
\end{figure*}

\begin{figure*}[h]
\subfloat[\label{fig:tomte_sec2}{Glitch labeled as \texttt{Tomte}. On the spectrogram with the longer time window we can see an overlap between a \texttt{Tomte} and \texttt{Scratchy} glitch.}]{%
\includegraphics[width=1\columnwidth]{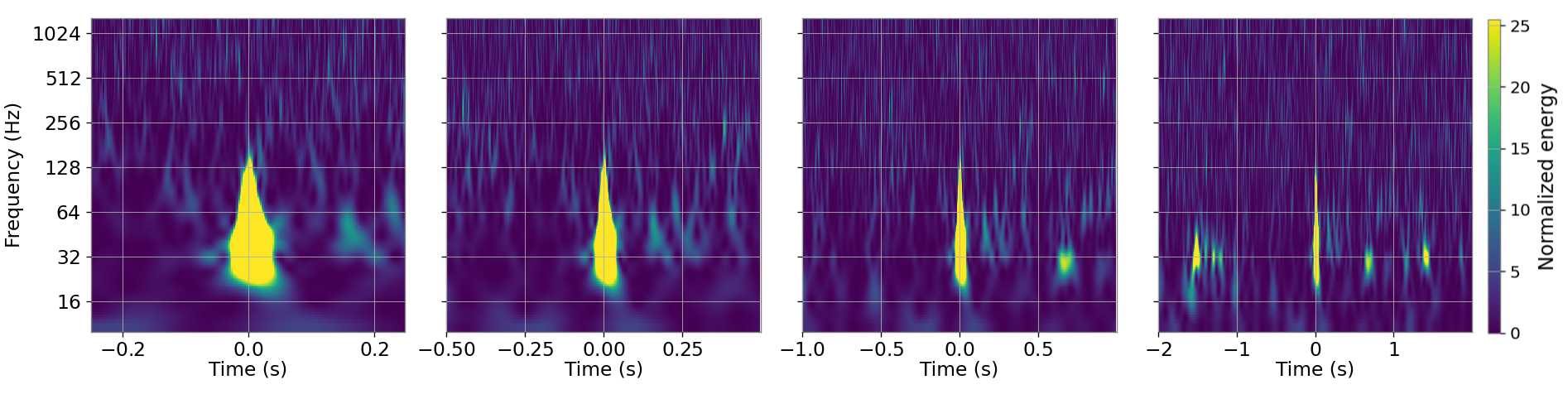}%
}\hfill
\subfloat[\label{fig:sclight_sec2}{Glitch labeled as \texttt{Scattered$\_$Light}. It is consistent with the \texttt{Fast$\_$Scattering} morphology, being a misclassification.}]{%
\includegraphics[width=1\columnwidth]{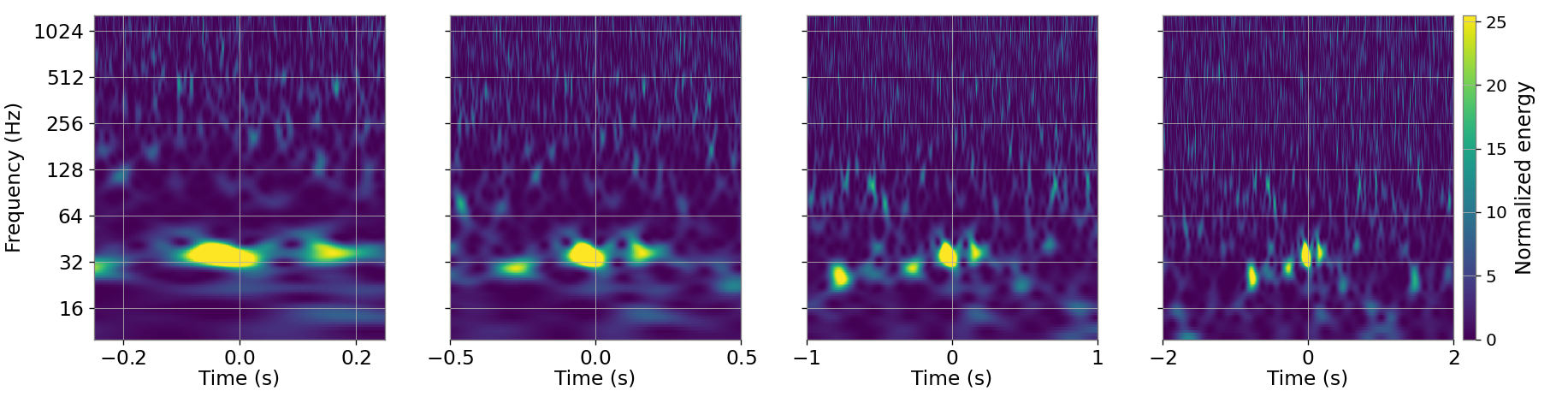}%
}\vfill
\subfloat[\label{fig:tomte2_sec2}{Glitch labelled as \texttt{Tomte}, overlapping with a smaller \texttt{Tomte}.}]{%
\includegraphics[width=1\columnwidth]{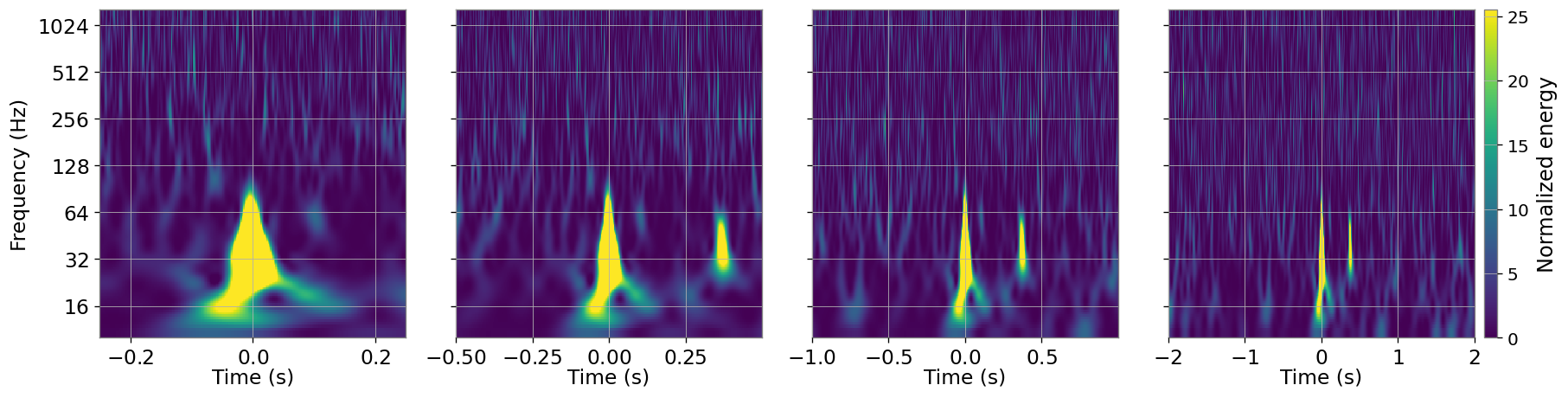}%
}\hfill
\subfloat[\label{fig:sclight2_sec2}{Glitch labeled as \texttt{Scattered$\_$Light}. On the left plot, an unknown morphology is overlapping at the beginning of the \texttt{Scattered$\_$Light}.}]{%
\includegraphics[width=1\columnwidth]{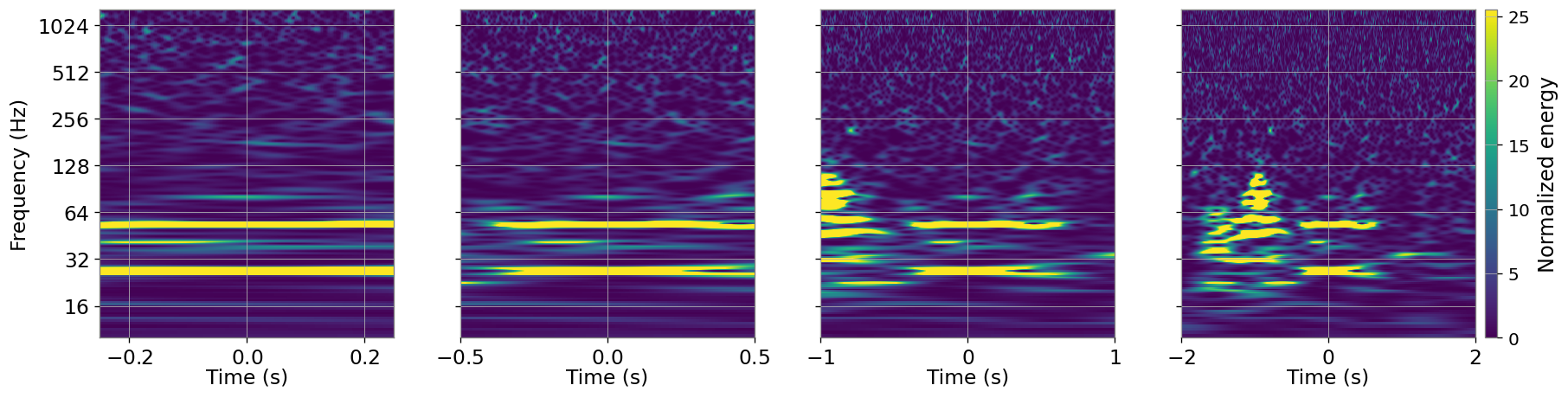}%
}
\caption{\textbf{Region 2:} Example of anomalous glitches in the embedded space.}
\label{fig:section2}
\end{figure*}

\begin{itemize}
    \item \textbf{Region 3} is a cluster with anomalous glitches from the 3 different glitch classes. Examples of anomalies are presented in Figs.~\ref{fig:whistle_sec3} and~\ref{fig:whistle2_sec3}, which are labelled as \texttt{Whistle} but have distinct morphologies that differ from any of the 22 Gravity Spy classes. Another anomalous glitch from this region is the misclassified glitch shown in Fig.~\ref{fig:tomte_sec3}, which was labelled as \texttt{Tomte} but has a morphology consistent with Fast Scattering. Another example of an anomalous glitch from this region presents as an overlap as shown in Fig.~\ref{fig:sclight_sec3}. The glitch was labelled as \texttt{Scattered$\_$Light} but seems to be a \texttt{Tomte} overlapping with an unknown morphology.
\end{itemize}

\begin{figure*}[h]
\subfloat[\label{fig:whistle_sec3}{Glitch labeled as \texttt{Whistle}, but it was not observed on other Gravity Spy's classes, so it could be a novel morphology.}]{%
\includegraphics[width=1\columnwidth]{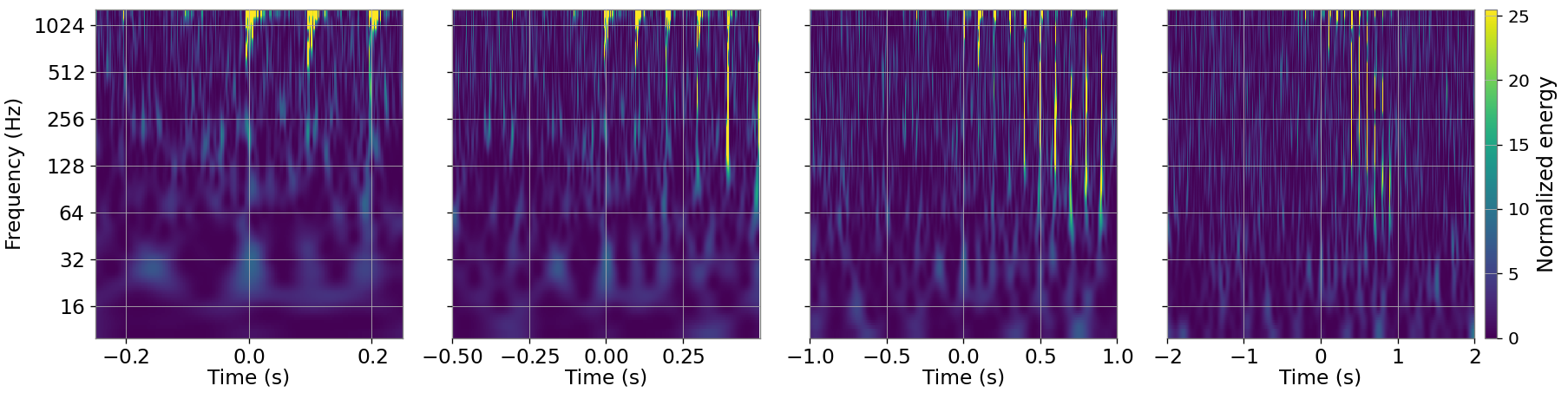}%
}\hfill
\subfloat[\label{fig:tomte_sec3}{Glitch labeled as \texttt{Tomte}, but its shape is consistent with \texttt{Fast$\_$Scattering}, being a misclassification.}]{%
\includegraphics[width=1\columnwidth]{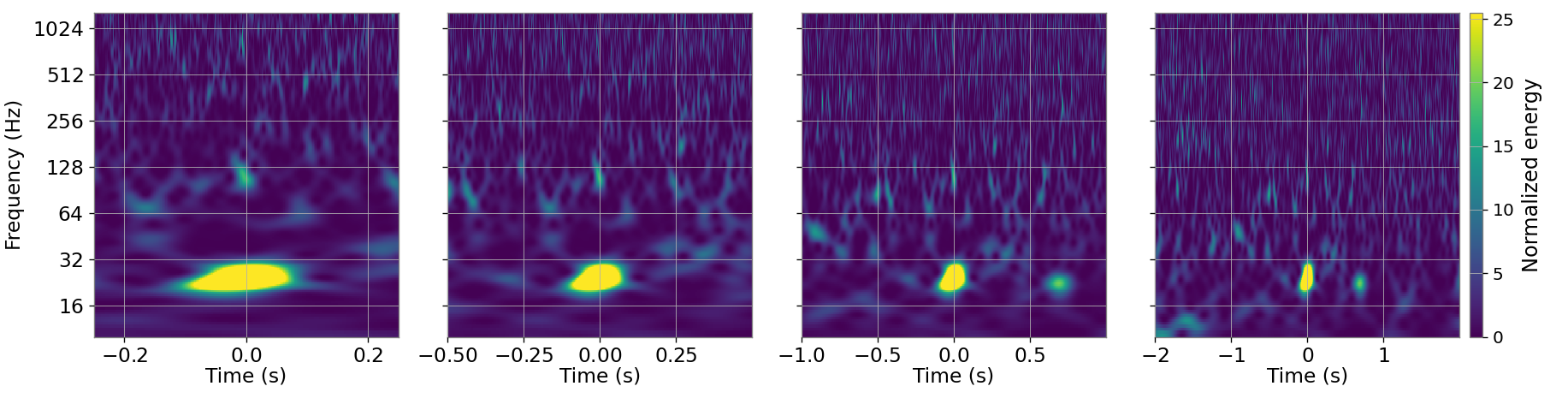}%
}\vfill
\subfloat[\label{fig:sclight_sec3}{Glitch labeled as \texttt{Scattered$\_$Light}, but its shape is consistent an overlap between a \texttt{Tomte} and an unknown morphology.}]{%
\includegraphics[width=1\columnwidth]{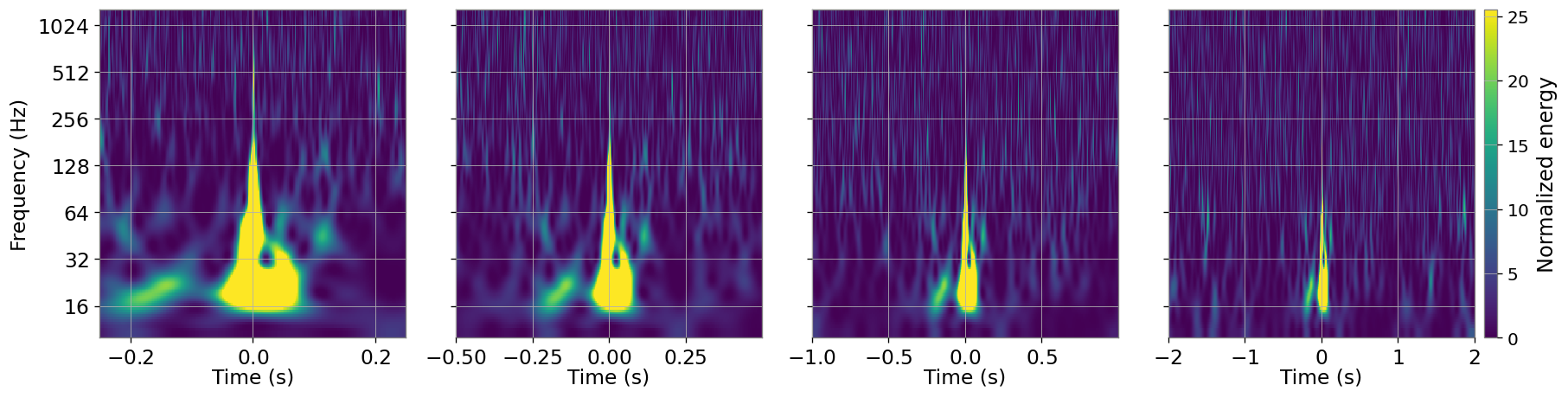}%
}\hfill
\subfloat[\label{fig:whistle2_sec3}{Glitch labeled as \texttt{Whistle}. The observed shape is inconsistent with any Gravity Spy classes, constituting an anomalous glitch morphology. }]{%
\includegraphics[width=1\columnwidth]{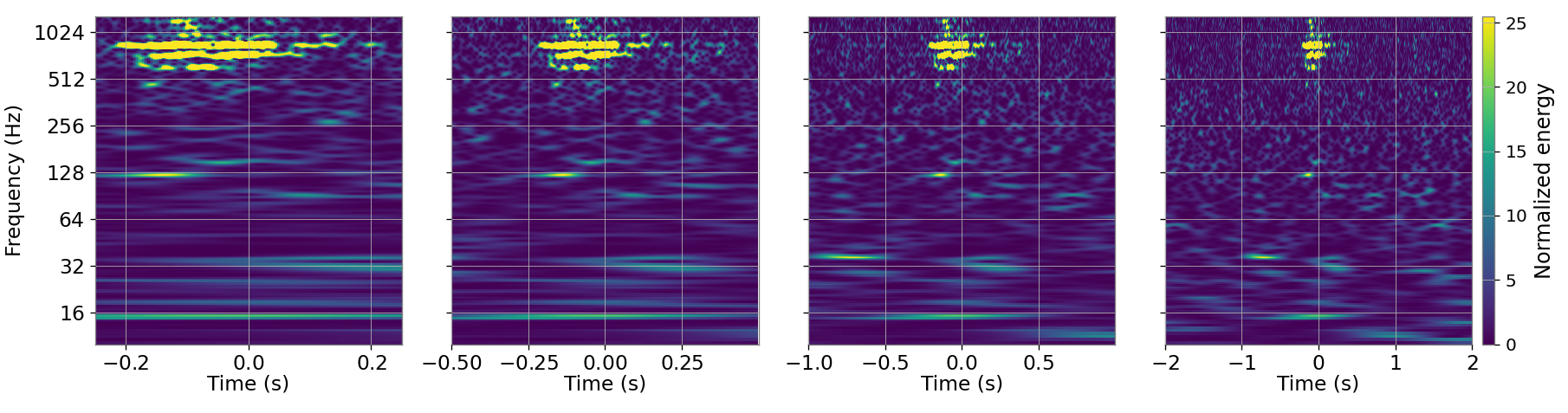}%
}
\caption{\textbf{Region 3:} Example of anomalous glitches in the embedded space.}
\label{fig:section3}
\end{figure*}

\begin{itemize}
    \item \textbf{Region 4} corresponds to the main \texttt{Scattered$\_$Light} cluster. In this region, there is a high presence of \texttt{Tomte} labels, which could indicate that both physical processes are related. In Fig.~\ref{fig:tomte1_sec4}, we present a glitch from this region that was labelled as \texttt{Tomte} but is consistent with the \texttt{Koi$\_$Fish} class, while \ref{fig:tomte2_sec4} was also labelled as a \texttt{Tomte} but seems to be an overlap between \texttt{Tomte} and Scratchy.
\end{itemize}

\begin{figure*}[h]
\subfloat[\label{fig:tomte1_sec4}{Glitch labeled as \texttt{Tomte}, but consistent with \texttt{Koi$\_$Fish} class, being a misclassification.}]{%
\includegraphics[width=1\columnwidth]{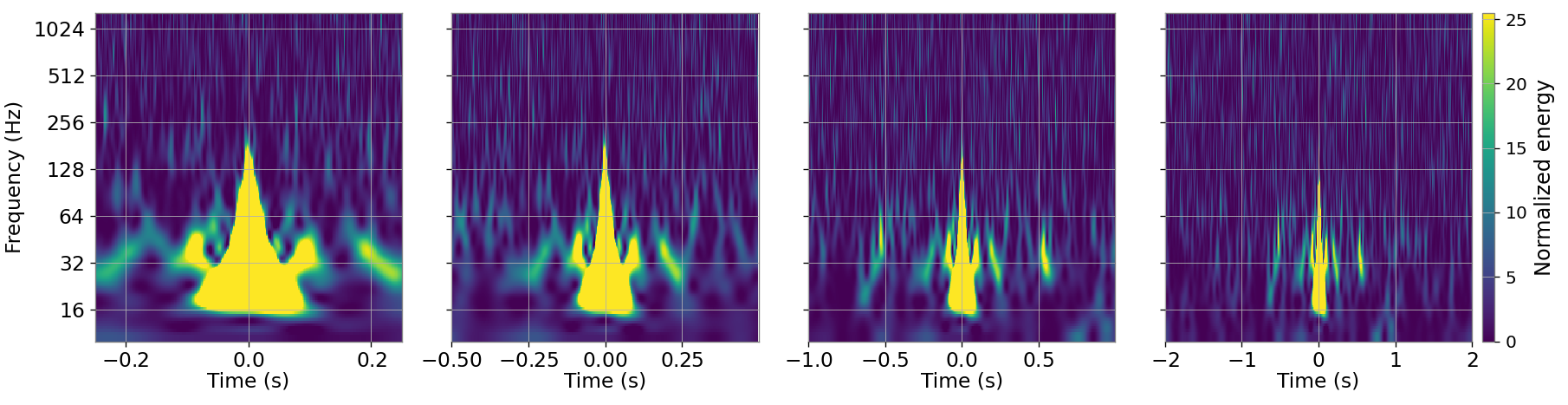}%
}\hfill
\subfloat[\label{fig:tomte2_sec4}{Glitch labeled as \texttt{Tomte}, but seems to be an overlap between a \texttt{Tomte} and a \texttt{Scratchy}, being an overlap.}]{%
\includegraphics[width=1\columnwidth]{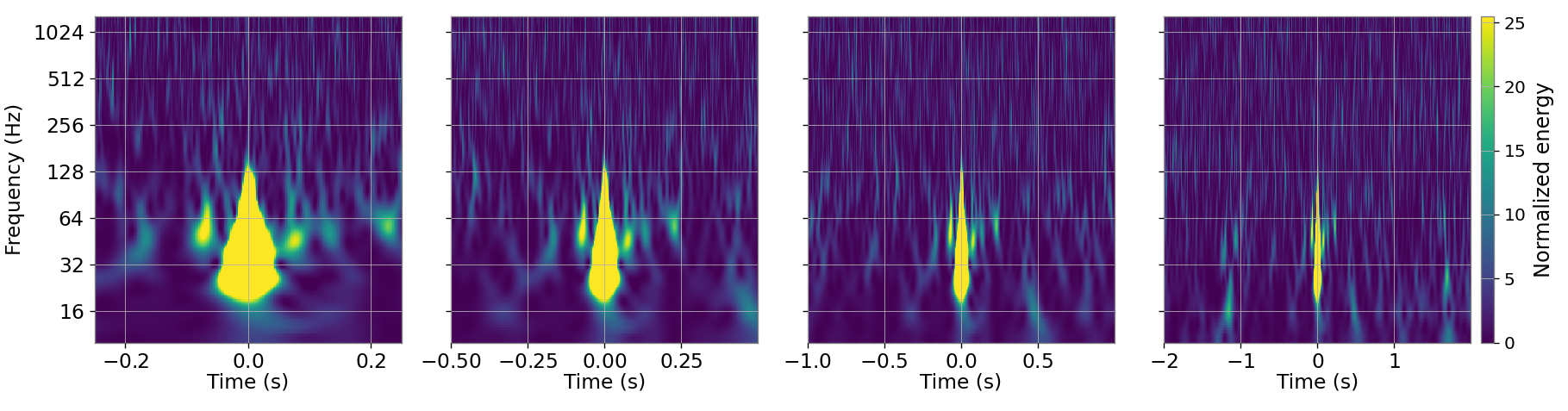}%
}
\caption{\textbf{Region 4:} Example of anomalous glitches in the embedded space.}
    \label{fig:section4}
\end{figure*}

\begin{itemize}
    \item \textbf{Region 5} corresponds to the main \texttt{Whistle} cluster, where we find some \texttt{Scattered$\_$Light} labels as well as a few \texttt{Tomte} labels. In Figs.~\ref{fig:tomte_sec5} and~\ref{fig:tomte2_sec5} we see \texttt{Tomte} glitches that appear to be overlapping with \texttt{Scratchy} glitches, in Fig.~\ref{fig:sclight_sec5} we see a glitch labelled as \texttt{Scattered$\_$Light} but could be a novel morphology, and in Fig.~\ref{fig:sclight2_sec5} we see a \texttt{Scratchy} glitch misclassified as \texttt{Scattered$\_$Light}.
\end{itemize}

\begin{figure*}[h]
\subfloat[\label{fig:tomte_sec5}{Glitch labeled as \texttt{Tomte}, but it could be an overlap between \texttt{Scratchy} and \texttt{Tomte}.}]{%
\includegraphics[width=1\columnwidth]{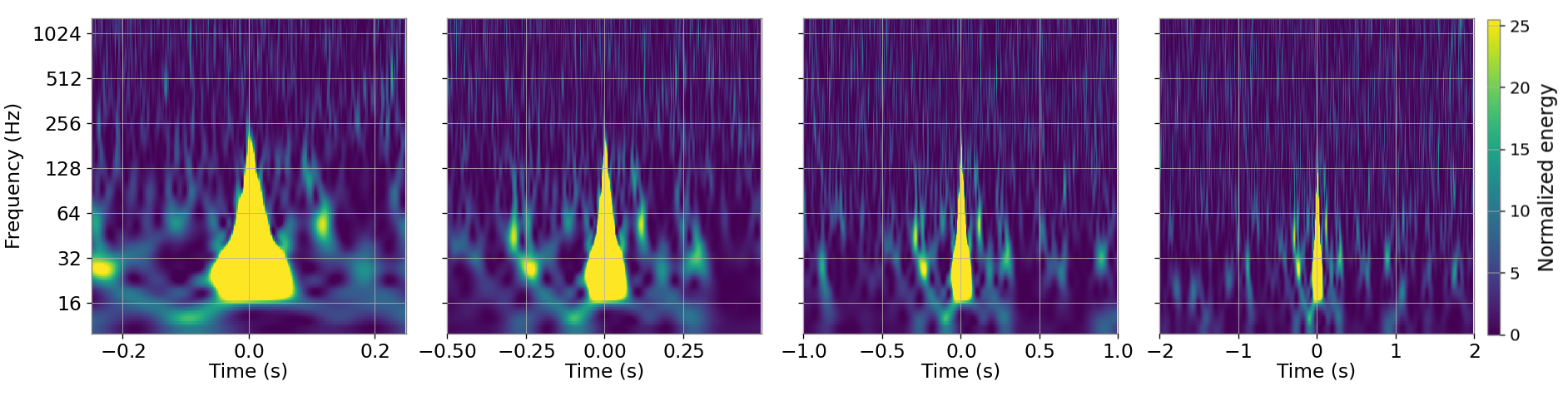}%
}\hfill
\subfloat[\label{fig:sclight_sec5}{Glitch labeled as \texttt{Scattered$\_$Light}, but it was not observed on other Gravity Spy's classes, so it could be a novel morphology.}]{%
\includegraphics[width=1\columnwidth]{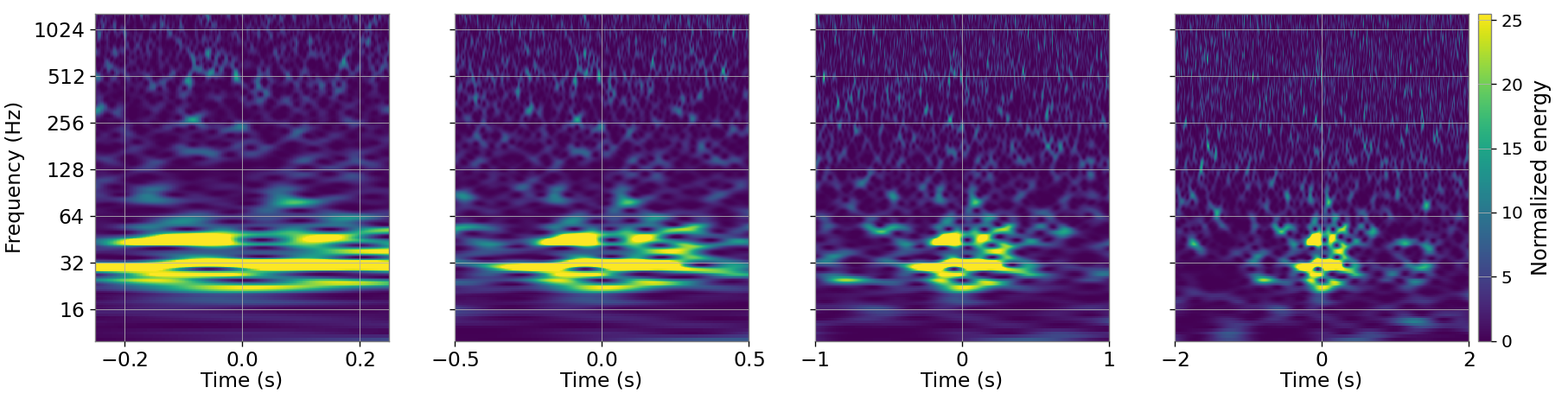}%
}\vfill
\subfloat[\label{fig:tomte2_sec5}{Glitch labeled as \texttt{Tomte}, but it could be an overlap between \texttt{Scratchy} and \texttt{Tomte}.}]{%
\includegraphics[width=1\columnwidth]{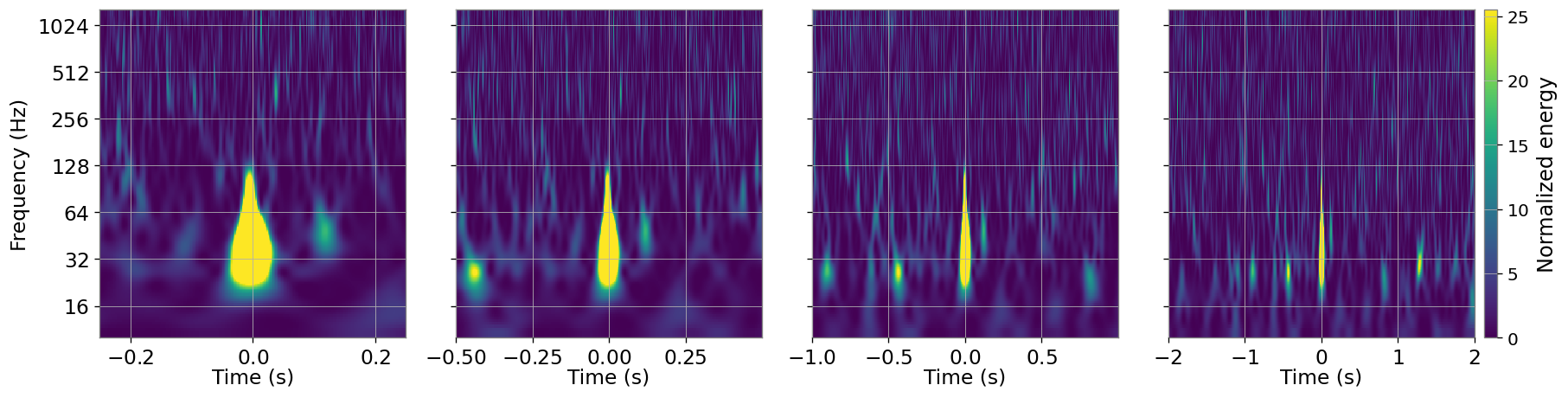}%
}\hfill
\subfloat[\label{fig:sclight2_sec5}{Glitch labeled as \texttt{Scattered$\_$Light}, but its morphology is consistent with \texttt{Scratchy}, being a misclassification.}]{%
\includegraphics[width=1\columnwidth]{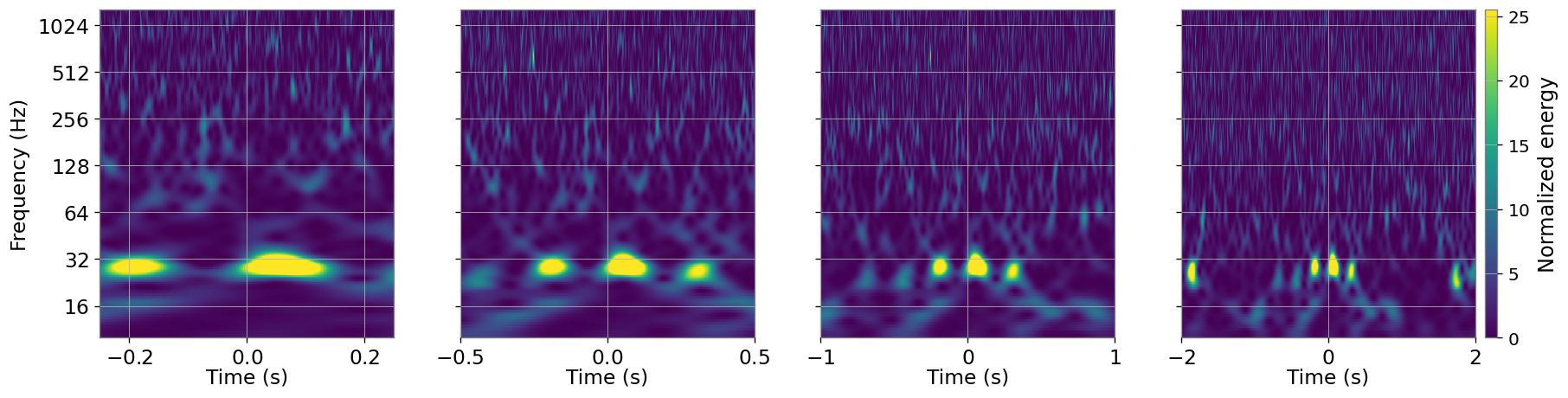}%
}
\caption{\textbf{Region 5:} Example of anomalous glitches in the embedded space.}
\label{fig:section5}
\end{figure*}

 {The outliers found at the outskirts of their clusters are visually and manually selected from the t-SNE representation in Fig.~\ref{fig:embeddedtsne}}, with the aim of automating the procedure in future works. After outliers have been selected, their spectrograms in $h(t)$ are visually inspected by comparing them to standard Gravity Spy morphologies. With this procedure, a total of 177 anomalies were found  {out of 2688 samples}, which implies $6.6 \%$ of the data. In particular, for each class, we found:

\begin{itemize}

\item \texttt{Whistle}: 49 anomalies were found, 45$\%$ are unknown morphologies, 28$\%$  are Gravity Spy misclassifications, and 27$\%$  are glitch overlaps.

\item \texttt{Tomte}:  57 anomalies were found, 32$\%$  are unknown morphologies, 21$\%$  are misclassifications, and 47$\%$  are glitch overlaps.

\item \texttt{Scattered$\_$Light}:  71 anomalies were found, 28$\%$  are unknown morphologies, 72$\%$  are misclassifications, and only one case of overlap is found.
\end{itemize}

After a visual inspection, we found that for \texttt{Whistle} most outliers constitute unknown morphologies, while for 
\texttt{Tomte} most anomalies are due to overlaps,  {where} it is common that two \texttt{Tomte} happen simultaneously. For \texttt{Scattered$\_$Light}, most outliers correspond to misclassifications, since the other seven glitch classes happen at similar frequency intervals and duration periods, namely \texttt{Low$\_$Frequency$\_$Burst}, \texttt{Low$\_$Frequency$\_$Lines}, \texttt{Power$\_$Line}, \texttt{Scratchy}, \texttt{Air$\_$Compressor}, \texttt{Paired$\_$Doves} and \texttt{Fast$\_$Scattering}.

 {While the misclassification of glitches could be countered with the improvement of training strategies, data set construction or class definitions, the identification of anomalies arising from overlaps and novel morphologies would still be hampered by the strict class definitions from supervised methods. Therefore, unsupervised approaches, such as the one presented in this work, will improve the understanding of glitch populations for their subsequent mitigation.}

\section{Conclusions}
\label{sec:conclusions}

{In this paper we have performed an exploratory analysis of a reduced set of safe auxiliary channels  from  LIGO Livingston with FD-encoding  in the context of anomaly detection.
{The focus of this work is, on one hand, to explore the potential of this data representation in the context of glitch characterization, and on the other hand, to build a data-driven model  {to cluster glitches in an unsupervised way with direct information from the detector, finding anomalies that deviate from the general distribution of the data}.} 

{For this aim,} we first speeded up the FD calculation from a computational complexity of $\mathcal{O}(N^3)$ in \cite{cavaglia2022characterization} to $\mathcal{O}(N^2 \log{N})$, constructing the FD-encoded safe auxiliary channel data set. Afterwards, we implemented a periodic convolutional autoencoder to learn the local and global structure of the data, compressed in a lower-dimensional space, known as embedded space. The reconstruction errors of the output of the autoencoder were $\sim 98.8\%$ of glitches $< 0.002$, implying that the autoencoder was able to learn the general trend of the data.}

{{We can also observe the reliable compression of the autoencoder, using solely safe auxiliary channels, when we project the embedded space in a two-dimensional t-SNE. This t-SNE representation clusters the different classes in separate regions which are consistent with Gravity Spy's observation in the main detector strain, $h(t)$. Samples that deviate significantly from their closest cluster are considered outliers.  {Representing these outliers in $h(t)$, we observed novel morphologies that strongly deviated from the standard definitions of Gravity Spy.}}%These outliers were represented in $h(t)$ to visualize their distinct morphology with respect to the labelling of the average Gravity Spy morphology. } 

{This methodology has shown that the safe auxiliary channel in the FD-encoding acts  {as a complementary representation to the visualization of $h(t)$, used to characterize the noise of the detector and to identify glitches for their subsequent mitigation.} Furthermore, our algorithm is flexible and completely data-driven, capable of uncovering misclassifications, glitch overlaps and novel glitch morphologies.  {While our method is independent of supervised classification algorithms, we used Gravity Spy as a benchmark to quantify its performance}: in  {our  FD-encoded auxiliary channel data, constituted by 2688 times where glitches were present in $h(t)$}, we found a 6.6$\%$ of anomalies caused by unknown morphologies labelled as their closest glitch class, similar morphologies assigned the incorrect class or glitch overlaps being overlooked. 

{{Data-driven approaches, such as the one demonstrated in this work,}  {can unveil anomalies present in the data and reveal relations between glitch classes, allowing us to further understand the glitch population.} { {In future work, this approach will be extended to the general population of LIGO-Livingston and other interferometers to enhance the identification of glitches. Moreover}, we will provide an anomaly score to assess the significance of the outliers found by our algorithm, and explore a data fusion representation containing both the FD-encoded auxiliary channel data and the strain $h(t)$ in time-frequency representation,  {providing not only information about the physical process within the detector but also their impact on $h(t)$}. {Last but not least, we will investigate the correlation between safe auxiliary channels highlighted by our model  {and glitches appearing in $h(t)$, in the search of witness auxiliary channels with the goal of improving  glitch mitigation in GW searches}.}}

\section{Acknowledgements}
M.L. is supported by the research program of the Netherlands Organisation for Scientific Research (NWO). S.C is supported by the National Science Foundation under Grant No. PHY-2309332. M.C. is supported by the National Science Foundation Grants NSF PHY-2011334 and NSF PHY-2308693. The authors are grateful for computational resources provided by the LIGO Laboratory and supported by the National Science Foundation Grants No. PHY-0757058 and No. PHY-0823459. This material is based upon work supported by NSF’s LIGO Laboratory which is a major facility fully funded by the National Science Foundation.

\bibliography{references}
\end{document}